\documentclass[12pt]{article}
\usepackage{cite}

\usepackage{amsmath}

\usepackage{amssymb}

\usepackage{verbatim}

\usepackage{tikz}

\usepackage{amsthm}
\newtheorem{proposition}{Proposition}

\usepackage[T1]{fontenc}  
\usepackage[english]{babel}
\usepackage{setspace}

\usepackage[colorlinks=true,linktocpage=true,linkcolor=blue,citecolor=blue,urlcolor=blue]{hyperref}

\newcommand{\nn}{\nonumber}
\newcommand{\wlong}{w_0}

\DeclareMathAlphabet{\mathbbold}{U}{bbold}{m}{n}  
\newcommand{\id}{\mathbbold{1}}

\begin{document}
\selectlanguage{english}
\allowdisplaybreaks[1]
\renewcommand{\thefootnote}{\fnsymbol{footnote}}
\numberwithin{equation}{section}
\def\corr{$\spadesuit \, $}
\def\trefle{$ \, $}
\def\kscorr{$\diamondsuit \, $}
\begin{titlepage}
\begin{flushright}
\
\vskip -2.5cm
{\small AEI-2013-271}\\
\vskip 2cm
\end{flushright}
\begin{center}
{\Large \bf
Fourier expansions of Kac--Moody Eisenstein\\[2mm]
series and degenerate Whittaker vectors}

\lineskip .75em
\vskip15mm
\normalsize
{\large  Philipp Fleig${}^{1,2}$, Axel Kleinschmidt${}^{1,3}$ and Daniel Persson${}^4$}

\vskip 1 em
${}^1${\it Max-Planck-Institut f\"{u}r Gravitationsphysik (Albert-Einstein-Institut)\\
Am M\"{u}hlenberg 1, DE-14476 Potsdam, Germany}
\vskip 1 em
${}^2${\it Freie Universit\"at Berlin, Institut f\"ur Theoretische Physik,\\
Arnimallee 14, DE-14195 Berlin, Germany}
\vskip 1 em
${}^3${\it International Solvay Institutes\\
ULB-Campus Plaine CP231, BE-1050 Brussels, Belgium}
\vskip 1 em
${}^4${\it Fundamental Physics, Chalmers University of Technology \\
412\,96 Gothenburg, Sweden}

\vskip10mm

\end{center}

\begin{abstract}
{ \footnotesize \noindent Motivated by string theory scattering amplitudes that are invariant under a discrete U-duality, we study Fourier coefficients of Eisenstein series on Kac--Moody groups.  In particular, we analyse the Eisenstein series on $E_9(\mathbb{R})$, $E_{10}(\mathbb{R})$ and $E_{11}(\mathbb{R})$ corresponding to certain degenerate principal series at the values $s=3/2$ and $s=5/2$ that were studied in~\cite{FK2012}. We show that these Eisenstein series have very simple Fourier coefficients as expected for their role as supersymmetric contributions to the higher derivative  couplings $\mathcal{R}^4$ and $\partial^{4}\mathcal{R}^4$ coming from 1/2-BPS and 1/4-BPS instantons, respectively. 
This suggests that there exist minimal and next-to-minimal unipotent automorphic representations of the associated Kac--Moody groups to which these special Eisenstein series are attached. We also provide complete explicit expressions for degenerate Whittaker vectors of minimal Eisenstein series on $E_6(\mathbb{R})$, $E_7(\mathbb{R})$ and $E_8(\mathbb{R})$ that have not appeared in the literature before.
 }
\end{abstract}

\end{titlepage}
%%%%%%%%%%%%%%%%%
\renewcommand{\thefootnote}{\arabic{footnote}}
\setcounter{footnote}{0}

%%%%%%%%%%%%%%%%%%%%%%%%%%%%%%%%%%%%%%%%%%%%%%%%%%%
\tableofcontents

\section{Introduction and Summary}

This paper concerns the analysis of Fourier coefficients of certain automorphic forms on infinite-dimensional Kac--Moody groups. As such, our results are of a mathematical nature, although our motivations come from the study of scattering amplitudes in string theory. 
In this introduction we begin in section \ref{sec_background} by setting the stage with  some general background and motivation, providing both the physical and the mathematical perspective on the subject of Eisenstein series and their Fourier expansions. We then describe some of our mathematical results in more detail in section  \ref{sec_summary}. Finally, section \ref{sec_outline} contains a short outline of the main text of the paper.

\subsection{Background}
\label{sec_background}
Automorphic forms are certain functions on Lie groups $G(\mathbb{R})$ with invariance properties under subgroups of the discrete $G(\mathbb{Z})$ group. Their importance in mathematics cannot be overstated: by now the theory of automorphic forms is a vast subject that touches upon several areas, ranging from number theory and geometry to representation theory, and beyond (see \cite{GelbartLanglands,KnappLanglandsProgram,FrenkelLanglands} for reviews). This web of interrelations between various fields is often referred to as the Langlands program, due to its origin in a series of far-reaching conjectures outlined by Langlands in \cite{LanglandsWeilLetter,LanglandsProb}. Much of the interesting information carried by an automorphic form is captured by its Fourier coefficients. In the classical case of $G(\mathbb{R})=SL(2, \mathbb{R})$ these coefficients often carry a wealth of arithmetic information, such as eigenvalues of Hecke operators and the counting of points on elliptic curves (see, e.g., \cite{Milne}). More generally, Fourier coefficients of automorphic forms on higher rank Lie groups  are closely linked with representation theory; for example, these coefficients play a key role in the transfer of automorphic forms from one group $G$ to another group $\tilde{G}$, a process known as ``functoriality'' --  one of the cornerstones of the Langlands program (see, e.g., \cite{ArthurFunctoriality,Cogdell}). 

Automorphic forms also appear naturally in physics, most notably in string theory. When supergravity is formulated on a (ten-dimensional) spacetime manifold which includes a compact submanifold $X$, the theory has a large moduli space of solutions. For certain choices of $X$ this moduli space is described by a coset space $G(\mathbb{R})/K(G)$, where the Lie group $G(\mathbb{R})$ is a continuous global symmetry and $K(G)$ is its maximal compact subgroup (corresponding to a local symmetry). When passing to string theory, the continuous symmetry $G(\mathbb{R})$ is broken to a discrete subgroup $G(\mathbb{Z})$ called ``U-duality'' group in physics~\cite{HullTownsend,ObersUdualMTheory}, and for the maximally supersymmetric cases the moduli space becomes the arithmetic double quotient $G(\mathbb{Z})\backslash G(\mathbb{R})/K(G)$. 

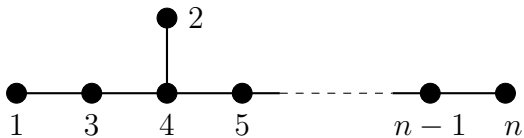
\begin{figure}[t]
\centering
\begin{tikzpicture}
[place/.style={circle,draw=black,fill=black,
inner sep=0pt,minimum size=7.5}]
\draw [thick] (-1,0) -- (-1,1);
\draw [thick] (-3,0) -- (0.5,0);
\draw [dashed] (0.5,0)--(2,0);
\draw [thick] (2,0) -- (3.5,0);
\node at (-3,0) [place,label=below:$1$] {};
\node at (-2,0) [place,label=below:$3$] {};
\node at (-1,0) [place,label=below:$4$] {};
\node at (0,0) [place,label=below:$5$] {};
\node at (2.5,0) [place,label=below:$n-1$] {};
\node at (3.5,0) [place,label=below:$\phantom{1}n$] {};
\node at (-1,1) [place,label=right:$2$] {};
\end{tikzpicture} 
\caption{\it
Dynkin diagram for $E_n$ with Bourbaki labelling. \label{fig:dynkin}}
\end{figure}

Physical observables, such as scattering amplitudes, must respect the symmetry and are therefore given by functions on $G(\mathbb{Z})\backslash G(\mathbb{R})/K(G)$; to wit, automorphic forms (see \cite{GutperleGreen,Green:1997di,Pioline:1998mn,GreenDualHighDeriv,ObersESeries,ObersStringThresh,Basu:2006cs,Basu:2007ru,Bao:2007er,Bao:2007fx,Pioline:2009qt,Bao:2009fg,Bao:2010cc,Green:2010wi,Gubay:2010nd,Lambert:2010pj,Lambert:2006ny} for a sample of the vast literature on the subject). Of particular interest for us is the case when $X=T^d$, the $d$-dimensional torus for $d=0, \dots, 10$. In this situation the relevant symmetry groups $G(\mathbb{R})$ fall into the series of  $E_{d+1}$-groups listed in Table~\ref{Udualitygroups} with  Dynkin diagrams  displayed in Figure~\ref{fig:dynkin}. In particular, for $d=5,6,7$ we have the exceptional Lie groups $E_6, E_7, E_8$, while for $d=8,9,10$ one conjecturally obtains discrete symmetries described by the affine, hyperbolic and Lorentzian Kac--Moody groups $E_9, E_{10}, E_{11}$, respectively \cite{HullTownsend}. 

\begin{table}[t!]
\begin{center}
\begin{tabular}{ | c || c  c  c  | }
  \hline                       
  $D/d$ & $E_{d+1}(\mathbb{R})$ & $K(E_{d+1})$ & $E_{d+1}(\mathbb{Z})$ \\ \hline \hline
 $10/0$ & $SL(2,\mathbb{R})$ & $SO(2)$ & $SL(2,\mathbb{Z})$ \\ \hline
 $9/1$ & $\mathbb{R}^+\times SL(2,\mathbb{R})$ & $SO(2)$ & $ SL(2,\mathbb{Z})$ \\ \hline
 $8/2$ & $SL(2,\mathbb{R})\times SL(3,\mathbb{R})$ & $SO(3)\times SO(2)$ & $SL(2,\mathbb{Z})\times SL(3,\mathbb{Z})$ \\ \hline
 $7/3$ & $SL(5,\mathbb{R})$ & $SO(5)$ & $SL(5,\mathbb{Z})$  \\ \hline
 $6/4$ & $SO(5,5,\mathbb{R})$ & $SO(5) \times SO(5)$ & $SO(5,5,\mathbb{Z})$ \\ \hline
 $5/5$ & $E_{6}(\mathbb{R})$ & $USp(8)$ & $E_{6}(\mathbb{Z})$  \\ \hline
 $4/6$ & $E_{7}(\mathbb{R})$ & $SU(8)/\mathbb{Z}_2$ & $E_{7}(\mathbb{Z})$ \\ \hline
 $3/7$ & $E_{8}(\mathbb{R})$ & $Spin(16)/\mathbb{Z}_2$ & $E_{8}(\mathbb{Z})$ \\ \hline
 $2/8$ & $E_{9}(\mathbb{R})$ & $K(E_{9}(\mathbb{R}))$ & $E_{9}(\mathbb{Z})$ \\ \hline
 $1/9$ & $E_{10}(\mathbb{R})$ & $K(E_{10}(\mathbb{R}))$ & $E_{10}(\mathbb{Z})$ \\ \hline
  $0/10$ & $E_{11}(\mathbb{R})$ & $K(E_{11}(\mathbb{R}))$ & $E_{11}(\mathbb{Z})$ \\ \hline
\end{tabular}
\caption{\label{Udualitygroups}\it This is a list of all U-duality groups which appear when compactifying maximal type IIB string theory in ten dimensions on a torus $T^d$ down to $D=10-d$ dimensions. The leftmost column corresponds to the global continuous symmetry of the classical theory~\cite{Cremmer:1979up}, and the middle column lists the associated maximal compact subgroups (which are local symmetries). In the rightmost column we give the discrete subgroups that are expected to be preserved in the string theory~\cite{HullTownsend}. The last two rows are conjectural as are the discrete groups $E_8(\mathbb{Z})$ and $E_9(\mathbb{Z})$.  }
\end{center}
\end{table}

{}From a physics perspective the Fourier coefficients of these automorphic forms capture quantum corrections to classical observables: the zeroth Fourier coefficient (or constant term) captures perturbative contributions, while the remaining Fourier coefficients capture non-perturbative (instanton) contributions \cite{GutperleGreen}. In order to determine which instanton configurations  contribute to a given observable it is therefore vital to know which of the Fourier coefficients are non-vanishing. As alluded to above, this question is also of major interest for mathematicians, and the key to answering it lies in the representation theory of the Lie group $G(\mathbb{R})$ \cite{GRS}. It is currently too ambitious to try and address this in the context of general amplitudes in string theory; instead we focus on a certain restricted class which is constrained by supersymmetry -- these are also called ``BPS-protected'' amplitudes. These types of amplitudes are characterized by having very few  quantum corrections, both at the perturbative and  the non-perturbative level. Put differently, the associated automorphic form must have very few non-vanishing Fourier coefficients. This is somewhat similar to the classical case of holomorphic modular forms $f(\tau)$ on the complex upper-half plane $\mathbb{H}$: holomorphicity then requires that the Fourier coefficients $a(n)$ in the $q=e^{2\pi i \tau}$-expansion $f(\tau)=\sum_{n\in \mathbb{Z}} a(n) q^{n}$ all vanish for $n<0$. Hence the BPS-condition can be thought of as a representation-theoretic generalization of this holomorphicity condition to automorphic forms on higher rank Lie groups. 

Recently there has been quite some progress  in understanding the relevant automorphic forms for higher derivative BPS-protected amplitudes in toroidal compactifications of maximal supergravity. Here we are mainly interested in curvature corrections to Einstein gravity in the low-energy effective action which are of the type $ \mathcal{R}^4$, $\partial^4\mathcal{R}^4$, where $\mathcal{R}^4$ is a specific contraction  of four Riemann tensors (see, e.g., \cite{Gross:1986iv}). These correspond to four-graviton superstring scattering in type IIB string theory compactified on a torus $T^d$. The associated corrections to the effective action are of the schematic forms 
\begin{align}
\int d^{10-d}x \sqrt{G} f^{(0)}_{E_{d+1}}(g) \mathcal{R}^4, \qquad \quad \int d^{10-d}x \sqrt{G} f^{(4)}_{E_{d+1}}(g)\partial^{4}\mathcal{R}^4, 
\label{corrections}
\end{align}
where $f_{E_{d+1}}^{(0)}$ and $f^{(4)}_{E_{d+1}}$ are functions on $E_{d+1}(\mathbb{R})$ satisfying 
\begin{align}
f^{(0)}_{E_{d+1}}(\gamma g k)=f^{(0)}_{E_{d+1}}(g), \qquad \gamma\in E_{d+1}(\mathbb{Z}), g\in E_{d+1}(\mathbb{R}), k\in K(E_{d+1}), 
\end{align}
and similarly for $f^{(4)}_{E_{n}}$, with $n=d+1$. In addition to the invariance properties given above, the coefficients $f_{E_{n}}^{(0)}$ and $f_{E_{n}}^{(4)}$ are subject to a number of physical constraints: $(i)$ they should be eigenfunctions of the Laplacian on the symmetric space $E_{n}/K$ with a specific eigenvalue (listed for example in~\cite{Green:2010wi}); $(ii)$ they must have a specific leading order behaviour when expanded around one of the cusps in $E_{n}/K(E_n)$. The first constraint arises from supersymmetry \cite{Green:1998by}, while the second is needed in order to match with known perturbative results. Hence, the coefficients are the sought-after automorphic forms that capture quantum corrections.

Let us illustrate this with an example. In the simplest case of $n=1$ ($d=0$) we have $E_1(\mathbb{R})=SL(2,\mathbb{R})$ and $K(E_1)=SO(2)$, such that the moduli space $SL(2,\mathbb{R})/K$ is isomorphic to the complex upper-half plane $\mathbb{H}$. This is parametrised by a complex variable $\tau=\tau_1+i\tau_2$ with $\tau_2>0$ such that the cusp at infinity corresponds to the limit $\tau_2\to \infty$. Physically, the variable $\tau_1$ is the axion of ten-dimensional IIB string theory, while $\tau_2=g_s^{-1}$ is the inverse string coupling. Hence, the limit $\tau_2\to \infty $ corresponds to weak coupling $g_s \to 0$. In the strict weak-coupling limit the leading order (tree-level) contribution to the $\mathcal{R}^4$ term is known to be $2\zeta(3)g_s^{-3/2}$ \cite{Gross:1986iv}. Hence, the automorphic form in front of $\mathcal{R}^4$ (in Einstein frame) must satisfy $f_{SL(2)}^{(0)}(\tau)\sim 2\zeta(3) \tau_2^{3/2}$ in the limit $\tau_2\to \infty$. This case was settled in the seminal paper by Green and Gutperle  \cite{GutperleGreen} where they showed that  $f_{SL(2)}^{(0)}(\tau)$ is the non-holomorphic Eisenstein series $E(s, \tau)=\sum_{(m,n)\neq (0,0)} \tau_2^{s}|m+n\tau|^{-2s}$ evaluated at $s=3/2$.\footnote{In principle one could also have contributions from so called \emph{cusp forms}, which have zero constant term (perturbative term), but these were subsequently ruled out in \cite{Pioline:1998mn} using a combination of U-duality and supersymmetry arguments.}

In a series of works \cite{Kiritsis:1997em,ObersStringThresh,Basu:2007ru,Green:2010wi,Green:2010kv,Pioline:2010kb,GreenSmallRep} (see also \cite{MillerSahi} for closely related mathematical results) this result has been generalized to higher-dimensional tori $T^{d}$, and by now  the functions $f^{(0)}_{E_{d+1}}$ and $f^{(4)}_{E_{d+1}}$ have been determined for $d= 0,1, \dots,7$: they are given by certain  Eisenstein series (see eq. (\ref{CurvEis}) below, as well as section \ref{sec_Eisenstein} for more details) attached to special automorphic representations of $E_{d+1}(\mathbb{R})$, whose Fourier coefficients have precisely the support expected from physics considerations.\footnote{For $d<5$ the functions  are actually linear combinations of Eisenstein series~\cite{Green:2010kv}. In the present article we are mainly interested in the case $d\geq 5$.} Specifically, the function $f^{(0)}_{E_{d+1}}$ should only receive contributions from (at least) 1/2 BPS-instantons, which forces severe restrictions on the number of non-vanishing Fourier coefficients. Following earlier work \cite{GRS}, it was shown in \cite{Pioline:2010kb,GreenSmallRep} that this condition is precisely satisfied for automorphic forms attached to the minimal unitary representation of the group $E_{d+1}(\mathbb{R})$ \cite{MR0404366,MR1159103,Gunaydin:2001bt}. Here, ``minimal'' refers to the fact that this representation is of smallest (non-trivial) functional dimension\footnote{Unitary representations of non-compact Lie groups $G(\mathbb{R})$ are infinite-dimensional, but one can still attach to them a notion of ``size'', called the functional (or Gelfand-Kirillov) dimension. This is defined as the smallest number of variables on which the associated functions can depend on. For example, the space $L^2(\mathbb{R}^n)$ of square-integrable functions on $\mathbb{R}^n$ has functional dimension $n$ in this sense.}. 

These results imply that  the automorphic form  $f^{(0)}_{E_{n}}(g)$ corresponds  to a special point in the parameter space of  Eisenstein series on $E_{n}(\mathbb{R})$. Generically these Eisenstein series depend on $n=\mathrm{rk}\,  E_{n}$ complex parameters $(s_1, \dots, s_n)$ and $f^{(0)}_{E_{n}}$ arises at the point $s=3/2$ on a certain complex one-dimensional  locus in $\mathbb{C}^n$, parametrised by $s$. This specialization is responsible for the vanishing of most of the Fourier coefficients which exist in the generic case. For example, according to work of Langlands \cite{LanglandsEP}, a generic Eisenstein series on a finite-dimensional simple Lie group $G(\mathbb{R})$ has the same number of constant terms as the order of the Weyl group of the associated Lie algebra $\mathfrak{g}$. This would give far too many perturbative contributions to match the coefficient of the $\mathcal{R}^4$-term, but when restricting to $s=3/2$ there are many cancellations in the general formula for the constant term and the number of constant terms precisely reduces to the known perturbative contributions \cite{Green:2010kv}. Mathematically, $s=3/2$ is precisely the value for which the Eisenstein series is attached to the minimal automorphic representation \cite{GRS}\footnote{In fact, this corresponds to $s=1/4$ in the normalisation of \cite{GRS}; see the conversion table on page 58 of \cite{GreenSmallRep}.}, thus explaining the collapse in the number of Fourier coefficients. 

A similar phenomenon occurs for the term involving $\partial^4\mathcal{R}^4$: its coefficient $f_{E_{n}}^{(4)}(g)$ corresponds to the value $s=5/2$ of the same generic Eisenstein series; this is then attached to the so called ``next-to-minimal'' representation of $E_{n}(\mathbb{R})$ \cite{Pioline:2010kb,GreenSmallRep}. Physically, this is a manifestation of the fact that the term $\partial^4\mathcal{R}^4$ in the effective action receives quantum corrections only from 1/2 and 1/4 BPS-states.

The situation becomes considerably more complicated when the torus $T^{d}$ has dimension eight or higher. In this situation we move into the realm of symmetry groups $E_{d+1}$ which are infinite-dimensional Kac--Moody groups. Garland has pioneered the study of Eisenstein series on affine Kac--Moody groups and has in particular analysed their convergence properties \cite{GarlConvergence}. Moreover, he also proved \cite{GarlLoop} a constant term  formula which generalizes the formula of Langlands for finite-dimensional Lie groups. For affine groups this formula involves a sum over the affine Weyl group, and has thereby infinitely many contributions. This is in stark contrast with what we expect from the associated physical functions $f_{E_n}^{(0)}(g)$ and $f_{E_n}^{(4)}(g)$, which should still  only have a finite number of perturbative contributions for reasons of supersymmetry. This puzzle was resolved in the paper \cite{FK2012} (see also \cite{Fleig:2012zc} for a short account) where it was shown that for the same special points in the parameter space, in particular for $s=3/2$ and $s=5/2$, the infinite number of constant terms in fact collapse to a finite number, matching  the expectations from physics. This result was also shown to hold for the formal generalization of Eisenstein series to $E_{10}(\mathbb{R})$ and $E_{11}(\mathbb{R})$. 

Motivated by these developments, in this paper we initiate a study of the non-constant Fourier coefficients of Eisenstein series on Kac--Moody groups (see also \cite{Liu,2013arXiv1306.3280C,Garland:2013lza,Bao:2013fga} for related work), and show that a similar collapse happens also here when specializing to $s=3/2$ and $s=5/2$ for  the groups $E_9(\mathbb{R})$ and $ E_{10}(\mathbb{R})$. Mathematically this is achieved by proving a formula for certain degenerate Whittaker vectors on $E_{n}(\mathbb{R})$ and show that upon restricting the parameters most of the terms in this formula vanish, effectively reducing the result to linear combinations of Whittaker vectors on $SL(2,\mathbb{R})$ or $SL(2,\mathbb{R})\times SL(2,\mathbb{R})$. 
As a by-product we also use our formula to calculate the complete list of the non-vanishing Whittaker vectors  on the finite-dimensional exceptional groups $E_6(\mathbb{R})$, $E_7(\mathbb{R})$ and $ E_8(\mathbb{R})$ in the minimal representation. 

We shall now summarize our mathematical results in a little more detail. 

\subsection{Summary of results}
\label{sec_summary}

Let $\mathfrak{g}$ be a Kac--Moody algebra and $G$ its associated Kac--Moody group. We are interested in the case when $G=G(\mathbb{R})$ is a split real form and $G(\mathbb{Z})\subset G(\mathbb{R})$ is a discrete subgroup of Chevalley type. $G(\mathbb{R})$ has an Iwasawa decomposition $G=BK=NAK$ and we formally define a Langlands-Eisenstein series on $G(\mathbb{R})$ according to\footnote{In the literature one also encounters the definition in terms of a (constrained) sum over the so-called instanton charge lattice  with carefully implemented BPS conditions~\cite{Ferrara:1997ci,ObersStringThresh}.}
\begin{align}
\label{BorelEisenstein}
E(\chi, g)=\sum_{\gamma\in B(\mathbb{Z})\backslash G(\mathbb{Z})} \chi(\gamma g).
\end{align}
Here, $\chi : G(\mathbb{R})\to \mathbb{C}^*$ satisfies $\chi(nak)=\chi(a)$. The value of $\chi$ on the Cartan subgroup $A$ can be parametrised by a choice of complex weight $\lambda \in \mathfrak{h}^*\otimes \mathbb{C}$, where $\mathfrak{h}^*$ is the dual space of the Cartan subalgebra $\mathfrak{h}=\mathrm{Lie}(A)$. Using this parametrisation one then writes $\chi(a)=a^{\lambda +\rho}$, where $\rho$ is the Weyl vector and $\lambda$ itself is parametrised by $n$ complex parameters $(s_1,\ldots,s_n)$ (see section \ref{sec_Eisenstein} for more details). The function $\chi$ is trivial under left-action of $B(\mathbb{Z})$, i.e., $\chi(\gamma g) = \chi(g)$ for all $\gamma\in B(\mathbb{Z})$, explaining the quotient in (\ref{BorelEisenstein}). Using the relation $\chi(a)=a^{\lambda+\rho}$, we also often write $E(\lambda,g)=E(\chi,g)$. 

When $G$ is a finite-dimensional Lie group, absolute convergence of $E(\lambda, g)$ in (\ref{BorelEisenstein}) was proven by Godement~\cite{Godement,BorelBoulder} in a domain where $\Re(\lambda)$ is large enough (see below eq.~\eqref{EisR} for a more concise statement) and analytic continuation to almost all complex $\lambda$ was established by Langlands~\cite{Langlands}. When $G$ is an affine Kac--Moody group convergence was established by Garland~\cite{GarlConvergence} (in the affine case one must be more cautious with how $\chi$ depends on the derivation and also with the precise value of $a\in A$~\cite{GarlLoop,GarlConvergence} but we circumvent this subtlety here by working solely in the domain of absolute convergence). The problem of proving convergence for more general  classes of Kac--Moody groups is still open, except for the case of rank 2 hyperbolic  groups which was recently established in \cite{2013arXiv1306.3280C}. 

As alluded to in the previous section we are interested in the Fourier expansion of the Eisenstein series $E(\lambda, g)$. The existence of such an expansion follows from the translational invariance $E(\lambda, ng)=E(\lambda, g)$ for all $n\in N(\mathbb{Z})=N(\mathbb{R})\cap G(\mathbb{Z})$. In order to write the expansion it turns out to be very convenient to pass to the adelic viewpoint, in which we consider the adelic group $G(\mathbb{A})$ in place of $G(\mathbb{R})$ and its discrete subgroup of rational points $G(\mathbb{Q})$ in place of $G(\mathbb{Z})$ -- we then view $E(\lambda, g)$ as a function on $G(\mathbb{Q})\backslash G(\mathbb{A})$. See section \ref{sec_adelisation} for more details on the process of adelisation. 

Let $\psi : N(\mathbb{A})\to U(1)$ be any unitary character on $N(\mathbb{A})$ which is trivial on $N(\mathbb{Q})$. Such a character is necessarily trivial on the entire derived subgroup $[N,N]$ and so restricts to a character on the abelianisation $[N, N]\backslash N$. We can then write the general form of the Fourier expansion as follows 
\begin{align}
E(\lambda, g) = C(\lambda, a)+ \sum_{\psi\neq 1} W_\psi(\lambda, g) + \text{non-abelian terms}\,,
\label{Fourierstructure}
\end{align}
where $C(\lambda, a)$ is the constant term and the Fourier coefficient $W_\psi(\lambda, g)$ is given by the Whittaker integral
\begin{align}
\label{Wdefintro}
W_\psi(\lambda, g) = \int\limits_{N(\mathbb{Q})\backslash N(\mathbb{A})} E(\lambda, ng) \overline{\psi(n)} dn.
\end{align}
The constant term is given by the same formula for the trivial character $\psi=\id$. The non-abelian terms correspond to Fourier coefficients which have a non-trivial dependence on $[N, N]$; these will not concern us in this paper.\footnote{Some results on non-abelian terms can be found for example in~\cite{VT,Bump,Pioline:2009qt,Bao:2009fg}.}

The dimension of the abelianisation $[N,N]\backslash N$ equals the rank $n$ of the (simple) group $G(\mathbb{R})$ and we can parametrise $\psi$ by an $n$-plet of rational numbers $m_\alpha \in \mathbb{Q}$ for all simple roots $\alpha$. The trivial character $\psi=\id$ corresponds to $m_\alpha=0$ for all simple roots $\alpha$. Note that by construction the Whittaker vector satisfies $W_\psi(\lambda, nak)=\psi(n) W_\psi(\lambda, a)$ and so $W_\psi(\lambda,g)$ is determined by its values on $A$. The right-$K$-invariance of $W_\psi(\lambda,g)$ is sometimes encoded by referring to it as a \textit{spherical} Whittaker vector. The importance of Whittaker vectors is also highlighted by the fact that one can sometimes rewrite the full automorphic function in terms of its Whittaker vectors~\cite{MR546599,MR0348047}

It turns out that, in contrast to the case of finite-dimensional Lie groups,  for infinite-dimensional Kac--Moody groups the Fourier coefficient $W_\psi(\lambda, g)$  always vanishes when $\psi$ is generic, i.e., all $m_\alpha\neq 0$ (see section \ref{sec_WhittakerKM} and~\cite{Liu,2013arXiv1306.3280C}). It is therefore of particular interest to compute the coefficient $W_\psi(\lambda, g)$ for \textit{degenerate Whittaker vectors}. By this we mean Whittaker vectors $W_{\psi}(\lambda,g)$ associated with characters $\psi$ for which at least one of the $m_\alpha$ vanishes.

Motivated by this we prove the following formula for the degenerate Whittaker vector in section \ref{degfour}:\footnote{A similar formula for Whittaker vectors on finite-dimensional semi-simple Lie groups was proven previously in \cite{MR665496}.}

\begin{proposition}\label{prop1}
{\it When $\psi$ is a  degenerate unitary character on $N(\mathbb{A})$ and $\lambda$ in the Godement range such that the Eisenstein series~\eqref{BorelEisenstein} converges absolutely, the Whittaker vector $W_\psi(\lambda, a)$ is given by 
\begin{align} 
W_\psi(\lambda, a)= \sum_{w_c\wlong'\in\, \mathcal{C}_\psi} a^{(w_c\wlong')^{-1}\lambda+\rho} M(w_c^{-1},\lambda) W'_{\psi^a}(w_c^{-1}\lambda,\id).
\label{degWhit}
\end{align}
Here $W'_\psi$ is a Whittaker vector for the subgroup $G'\subset G$ such that $\psi$ is generic on the unipotent radical $N'\subset G'$; $\psi^{a}$ denotes a certain twisted version of $\psi$ defined in section \ref{sec_twist}. The set $\mathcal{C}_\psi\subset \mathcal{W}$ is the set of Weyl words (potentially) contributing to $W_\psi$; its explicit parametrisation in terms of the longest Weyl $\wlong'$ of $G'$ and special coset representatives $w_c$ is defined in section~\ref{sec_WeylWords}. Finally, the factor $M(w,\lambda)$ is the same product over 
ratios of completed zeta functions that appears in the constant term formula of Langlands~\cite{LanglandsEP} (see also~\cite{Moeglin}) and is given explicitly in eq.~(\ref{Mw}). %By the functional equation of $E(\lambda,g)$, the formula~\eqref{degWhit} extends to almost all complex $\lambda$.
}
\end{proposition}

In particular, for maximally degenerate characters $\psi$ (i.e. only one $m_\alpha\neq 0$), the Whittaker vector $W_\psi(2s\Lambda_{i_*}-\rho, a)$ reduces to a linear combination of (twisted) $SL(2, \mathbb{R})$ Whittaker vectors. We refer to these Whittaker vectors as vectors of type $A_1$. For degenerate characters $\psi$ that have two non-vanishing $m_\alpha$ on non-adjacent nodes with subgroup $G'=SL(2,\mathbb{R})\times SL(2,\mathbb{R})$,  one obtains the product of two $SL(2,\mathbb{R})$ Whittaker vectors. We refer to this case as type $A_1\times A_1$. These two are all the cases we will require in this paper.

Our proof of proposition~\ref{prop1} is strictly valid only in the finite-dimensional case. Garland's results on affine Eisenstein series~\cite{GarlConvergence,GarlLoop} strongly suggest that our manipulations in the proof are also valid in this case and we will in fact assume that proposition~\ref{prop1} is also valid for Eisenstein series on general Kac--Moody groups in the domain of absolute convergence. Since our main interest is in very special Eisenstein series that are thought to arise in physics, we do not attempt a general proof for (complete) Kac--Moody groups in this paper.

In the general Kac--Moody case, the sum over the Weyl elements in (\ref{degWhit}) is, however, infinite. As mentioned in the previous section, we are interested in special choices of the parameter $\lambda$ such that this collapses to a finite sum. To this end we restrict to Eisenstein series $E(\lambda, g)$ for which 
\begin{align}
\lambda =2s\Lambda_{i_*}-\rho,
\end{align}
where $s\in \mathbb{C}$ and $\Lambda_{i_*}$ denotes the fundamental weight associated with a single simple root $\alpha_{i_*}$. As will be explained in section \ref{sec_Eisenstein} the associated Eisenstein series $E(2s\Lambda_{i_*}-\rho, g)$ can be interpreted as an Eisenstein series induced from a character on the maximal parabolic subgroup $P_{i_*}\subset G$, instead of the Borel subgroup as was the case in (\ref{BorelEisenstein}). The automorphic forms $f^{(0)}_{E_{n}}$ and $f^{(4)}_{E_{n}}$ discussed in section \ref{sec_background} are given by the Eisenstein series  
\begin{align}
f^{(0)}_{E_{n}}(g)=E(3\Lambda_1-\rho, g) \qquad\text{and} \qquad f^{(4)}_{E_{n}}(g)=E(5\Lambda_1-\rho, g), 
\label{CurvEis}
\end{align}
for $g\in E_{n}(\mathbb{R})$. For these Eisenstein series the sum over characters $\psi$ in (\ref{Fourierstructure}) then corresponds to a sum over instanton charges. 

For these more special Eisenstein series we now have the following results:

\begin{proposition}\label{prop2}
\textit{For the special choices of $(s, \Lambda_{i_*})$ which are relevant for the $\mathcal{R}^4$ and $\partial^4 \mathcal{R}^4$ curvature corrections in the low-energy effective action, the infinite sum in (\ref{degWhit}) collapses to a {\em finite} sum over elements in the set $\mathcal{C}_{\lambda,\psi}$ defined in eq. (\ref{collapsingset}). 
(This proposition is proven in section~\ref{sec_KMWhittaker}.)}
\end{proposition}

\begin{proposition}\label{prop3}
\textit{For the $\mathcal{R}^4$ curvature corrections one has $(s, \Lambda_{i_*})=(3/2, \Lambda_1)$ (with labelling as in figure~\ref{fig:dynkin})~\cite{FK2012}. The Whittaker vectors for the corresponding Eisenstein series $E_9$, $E_{10}$ and $E_{11}$ are given explicitly in section \ref{sec_KMtables}. In particular, for these cases we show that they are given by sums over $A_1$-type Whittaker vectors only such that
\begin{align}
\sum_{\psi\neq 0} W_\psi(\lambda, na) =\sum_{\alpha\in \Pi} \sum_{\psi_\alpha} c_\alpha(a)W'_{\psi^{a}_\alpha}(\chi'_\alpha, \id)\psi_\alpha(n), 
\end{align} 
where $\psi_\alpha$ denotes a maximally degenerate character with only  $m_\alpha\neq 0$ and $c_\alpha(a)$ is a simple polynomial function on the Cartan torus; $\Pi$ denotes the set of simple roots. In other words, all the Whittaker vectors vanish except the ones associated with maximally degenerate characters. 
}
\end{proposition}

\begin{proposition}\label{prop4}
\textit{For the $\partial^4\mathcal{R}^4$ curvature corrections one has $(s, \Lambda_{i_*})=(5/2, \Lambda_1)$ (with labelling as in figure~\ref{fig:dynkin})~\cite{FK2012}. The Whittaker vectors for the corresponding Eisenstein series on $E_9$, $E_{10}$ and $E_{11}$ are given by finite sums of $A_1$- and $A_1\times A_1$-type Whittaker vectors only. A representative example for $E_{10}$ is given in section~\ref{sec:E1052}.
}
\end{proposition}

For completeness, we also calculate the complete list of degenerate Whittaker vectors for $(3/2, \Lambda_1)$, corresponding to the minimal representation, for the finite-dimensional exceptional groups $E_6, E_7, E_8$; these results are contained in Appendix \ref{sec_finite}. They are also of $A_1$-type only.

\subsection{Outline} 
\label{sec_outline}

This paper is organized as follows. In section \ref{sec_Eisenstein} we introduce some basic properties of Langlands-type Eisenstein series, and their adelic formulation. We also recall the structure of Langlands' constant term formula. In section \ref{sec_FourierExpansions} we discuss the general structure of Fourier expansions of Eisenstein series and the connection with Whittaker vectors. We discuss in particular conditions for when these Whittaker vectors are non-zero, observing that in the Kac--Moody case this can only happen when the character $\psi$ is degenerate. This leads naturally to our analysis of degenerate Whittaker vectors in section \ref{degfour}. We prove formula~(\ref{degWhit}) of proposition~\hyperref[prop1]{\ref*{prop1}}, expressing these degenerate Whittaker vectors as Weyl sums over generic Whittaker vectors of subgroups. In section \ref{sec_KMWhittaker} we then apply these results to show that the infinite Weyl sum collapses to a finite sum for certain special Eisenstein series of relevance in string theory, proving propositions~\hyperref[prop2]{\ref*{prop2}},~\hyperref[prop3]{\ref*{prop3}} and~\hyperref[prop4]{\ref*{prop4}}. We provide explicit and full results for the Kac--Moody groups $E_9$,  $E_{10}$ and $E_{11}$ for the $\mathcal{R}^4$ correction and representative examples for the $\partial^4\mathcal{R}^4$ correction for $E_{10}$ in section~\ref{sec_KMtables}. We conclude in section \ref{sec_conclusions} with some open problems. The paper also contains some appendices. In appendix \ref{sec_finite} we present explicitly all degenerate Whittaker vectors for the exceptional Lie groups $E_6$, $E_7$ and $E_8$ for the $\mathcal{R}^4$ correction, and in appendix \ref{sec_proof} we present an alternative proof of the formula (\ref{degWhit}) for degenerate Whittaker vectors. Appendix~\ref{param} contains some technical details on a parametrisation of certain roots that play a role in the proofs. Finally, appendix~\ref{app:e10dist} lists distinguished gradings of $E_{10}$ itself.

\section{Eisenstein series and their adelic formulation}
\label{sec_Eisenstein}

In this section we briefly give the definition and key properties of Eisenstein series on the real Lie group $G(\mathbb{R})$ and their adelisation. More information can be found for example in~\cite{LanglandsEisenstein,Deitmar,FGKP}.

\subsection{Definition of Eisenstein series}

Let $G=G(\mathbb{R})$ be a split real simple Lie group of rank $n$; $G$ can be either of finite or infinite dimension. The cases we have in mind are $G=E_{n}$ for $n=6,7,8,9,10,11\ldots$. We restrict $\mathrm{Lie}(G)$ to be simply-laced for simplicity. The Eisenstein series on the moduli space $G/K$ that is invariant under the U-duality group $G(\mathbb{Z})$ can be induced from a perturbative term as follows. 

The ``coupling constants''\footnote{This includes string coupling constant $g_s$ and the $d$ compactification radii of the torus $T^d$ such that there are $n=d+1$ coupling constants in total.} of the theory sit in the maximal split torus $A\subset G$. We suppose that a perturbative term is known and that it is of the form $v_1^{2s_1} v_2^{2s_2}\cdots v_n^{2s_n}$, where the $v_i$ label the coupling constants, and the constants $s_i$ parametrise the dependence of the perturbative term on the coupling constants. There will also be an overall numerical coefficient of the term that we can absorb in the normalisation of the automorphic function to be defined presently. An example would be the perturbative string tree level correction to four-graviton scattering associated with a term $\mathcal{R}^4$: As discussed in the introduction, this term is of the form $2\zeta(3) g_{\mathrm{s}}^{-3/2}$ (in Einstein frame)~\cite{Gross:1986iv} and the string coupling $g_{\mathrm{s}}$ is related to the $v_i$ in a polynomial way and the overall coefficient $2\zeta(3)$ is the overall normalisation that we will no longer discuss. Given such a perturbative term, we can construct its $G(\mathbb{Z})$-completion by summing over all its images. A more mathematical description of this leads directly to Langlands' Eisenstein series~\cite{LanglandsEisenstein,GutperleGreen}, as we now demonstrate.

 Let $\chi: G \to \mathbb{C}$ be a function that projects a group element onto the perturbative term, i.e. 
 \begin{align}
\chi(g) = v_1^{2s_1}\cdots v_n^{2s_n},
 \end{align} 
 where we parametrise the Cartan torus $A\subset G$ by
 \begin{align}
 a = \exp \left[ \log(v_1) h_1+\ldots +\log(v_n) h_n\right] = v_1^{h_1}\cdots v_n^{h_n},
\end{align}
where the $h_i$ are the standard Chevalley generators of the Cartan subalgebra (their Killing inner product matrix is the Cartan matrix). In other words, $\chi$ assigns to a point of the moduli space $G/K$ a perturbative term. Furthermore, the function $\chi$ satisfies 
\begin{align}
\chi( nak) =\chi(a) 
\end{align}
where $g=nak$ is the Iwasawa decomposition and we will, loosely speaking, refer to it as a quasi-character.  The function $\chi$ is invariant under discrete Borel elements $B(\mathbb{Z})$: $\chi(\gamma g) =\chi(g)$ for $\gamma\in B(\mathbb{Z})$. Physically, these transformations correspond to discrete large gauge transformations of the axions. We also allow the parameters $s_i$ to take complex values. It will sometimes be convenient to embed the Weyl group $\mathcal{W}=\mathcal{W}(G)$ in the maximal compact subgroup $K\subset G$, thus making $\chi$ right-invariant under Weyl transformations.

Starting from $\chi$ we define the Eisenstein series
\begin{align}
\label{EisR}
E(\chi,g) = \sum_{\gamma\in B(\mathbb{Z})\backslash G(\mathbb{Z})} \chi(\gamma g).
\end{align}
This is the sum over all $G(\mathbb{Z})$-images of the perturbative term where the $\chi$-stabilising $B(\mathbb{Z})$ transformations have been quotiented out in order not to overcount the sum. The convergence of this sum depends on the group $G$ and the values $s_i$, where convergence is guaranteed by the condition $\Re(s_i)>1$ for all $s_i$, known as Godement's criterion~\cite{BorelBoulder}. This convergence was proven for finite-dimensional groups in~\cite{Langlands,LanglandsEisenstein}, for affine groups in~\cite{GarlLoop,GarlConvergence} and argued for more general Kac--Moody groups in~\cite{FK2012,Garland:2013lza,Bao:2013fga}. In the finite-dimensional case the domain can be extended to almost all complex values of $s_i$ by means of a functional equation~\cite{Langlands}; for the Kac--Moody case convergence appears restricted to the Tits cone~\cite{GarlConvergence,Garland:2013lza} (see~\cite{Kac90} for the notion of Tits cone) and one also needs to restrict the domain of $a$~\cite{2013arXiv1306.3280C}.

It turns out to be convenient to parametrise the function $\chi$ alternatively by means of a weight $\lambda$ of $G$. This is done by writing
\begin{align}
\label{chilambda}
\chi(a) = a^{\lambda+\rho},
\end{align}
where $\rho$ is the Weyl vector of $G$ and therefore $\lambda= \sum_{i} (2s_i-1)\Lambda_i$ in terms of the fundamental weights $\Lambda_i$ satisfying $\Lambda_i(h_j) = \delta_{ij}$ and we write the Eisenstein series alternatively as
\begin{align}
E(\lambda,g) \equiv E(\chi,g).
\end{align}
The functional relation allowing extension of the domain is then~\cite{LanglandsEisenstein}
\begin{align}
E(\lambda,g) = M(w,\lambda) E(w\lambda,g),
\end{align}
where $w\in\mathcal{W}$ is an element of the Weyl group of $G$ and 
\begin{align}
\label{Mw}
M(w,\lambda) = \prod_{\alpha>0\,|\, w\alpha<0} \frac{\xi(\langle\lambda|\alpha\rangle)}
{\xi(1+\langle\lambda|\alpha\rangle)},
\end{align}
is an important numerical coefficient that involves a product over all positive roots $\alpha$ that are mapped to negative roots by $w$. The function $\xi(k)=\pi^{-k/2}\Gamma(k/2)\zeta(k)$ is the completed Riemann $\zeta$-function and the angled bracket denotes the canonical inner product on the space of weights. The factor $M(w,\lambda)$ enjoys the important multiplicative property
\begin{align}
\label{Mmult}
M(w\tilde{w},\lambda) = M(w,\tilde{w}\lambda) M(\tilde{w},\lambda).
\end{align}

The Eisenstein series that have played a role in string theory thus far have been of a special type. More precisely, they originate from very simple perturbative terms that lead to weights $\lambda$ of the form
\begin{align}
\label{maxlambda}
\lambda = 2s \Lambda_{i_*} -\rho,
\end{align}
where $i_*$ denotes a single node of the Dynkin diagram. Instead of depending on $n$ different parameters $s_i$, the resulting Eisenstein series depends only a single parameter $s$. This case is referred to as maximal parabolic Eisenstein series in~\cite{Green:2010kv} because the corresponding characters $\chi$ have an enlarged invariance under a maximal parabolic subgroup $P_{i_*}(\mathbb{Z})$ rather than just $B(\mathbb{Z})$. As indicated, the maximal parabolic is defined by the node $i_*$ that also enters the definition of the weight in (\ref{maxlambda}). The string tree level term $\chi(g)=g_s^{-3/2}$ is exactly of this form. 

\subsection{Adelisation}
\label{sec_adelisation}

When analysing Eisenstein series it is often convenient to not treat them as functions on the real Lie group $G=G(\mathbb{R})$ but instead consider them as functions on the group $G(\mathbb{A})$ over the (rational) adeles $\mathbb{A}$. The validity of this extended viewpoint is guaranteed by the strong approximation theorem~\cite{MR670072} and explained for example in~\cite{MR546598}. The advantage of this is that more mathematical tools are available when performing  operations on $E(\chi,g)$. 

The adeles $\mathbb{A}$ over $\mathbb{Q}$ are defined as 
\begin{align}
\mathbb{A} = \mathbb{R} \times \sideset{}{'}\prod_{p<\infty} \mathbb{Q}_p,
\end{align}
where $\mathbb{Q}_p$ are the $p$-adic numbers~(see for instance \cite{Deitmar,FGKP} for an introduction) and the product involves all inequivalent completions of the field $\mathbb{Q}$ of rational numbers. The prime on the product indicates that almost all elements in this infinite product are restricted to the $p$-adic integers $\mathbb{Z}_p$.

The adelic version of the Eisenstein series, with the definition of the character $\chi$ appropriately extended to $G(\mathbb{A})$, is
\begin{align}
\label{EisA}
 E(\chi,g_{\mathbb{A}}) = \sum_{\gamma\in B(\mathbb{Q})\backslash G(\mathbb{Q})} \chi(\gamma g_{\mathbb{A}}),
\end{align}
where the difference to (\ref{EisR}) is that now $g_{\mathbb{A}}\in G(\mathbb{A})$ and the sum is over the diagonally embedded discrete subgroup $G(\mathbb{Q})$. The link to (\ref{EisR}) is obtained by restricting the element $g_{\mathbb{A}}$ to lie solely in the real factor:
\begin{align}
g_{\mathbb{A}} = (g_{\mathbb{R}}, 1,1,\ldots), \qquad \quad g_\mathbb{R}\in G(\mathbb{R}).
\end{align}
Evaluating the adelic Eisenstein series for such $g_{\mathbb{A}}$ defines a function on the real group $G(\mathbb{R})$ and, by strong approximation, this function is equal to (\ref{EisR}) defined above. We will in the sequel drop the subscript on the group element as it will be clear from the context whether $g$ is in $G(\mathbb{A})$ or $G(\mathbb{R})$.

One cornerstone of the adelic treatment is that one can write the quantity $M(w,\lambda)$ of (\ref{Mw}) as
\begin{align}
\label{Mwint}
M(w^{-1},\lambda) = \int\limits_{w^{-1}B(\mathbb{Q})w\cap N(\mathbb{Q})\backslash N(\mathbb{A})} \chi(wn) dn.
\end{align}
This integral arises naturally when calculating the constant term of the Eisenstein series that is defined by
\begin{align}
C(\lambda,g) = \int\limits_{N(\mathbb{Q})\backslash N(\mathbb{A})} E(\chi, ng) dn
\end{align}
and this is a function solely on the Cartan torus. Using the Bruhat decomposition and the integral (\ref{Mwint}) one can demonstrate Langlands' constant term formula
\begin{align}
\label{LCF}
C(\lambda,a) = \sum_{w\in\mathcal{W}} M(w,\lambda) a^{w\lambda+\rho},
\end{align}
where the notation (\ref{chilambda}) was used. More information on the adelic treatment of Eisenstein series can be found in~\cite{FGKP}.

\section{Fourier expansion of Eisenstein series}
\label{sec_FourierExpansions}

In this section, we discuss the general structure of Fourier expansions of Eisenstein series $E(\chi,g)$ and set up some of our basic notation. Fourier expansions can be defined with respect to arbitrary unipotent radicals $U$ of the group $G$ that the Eisenstein series is defined on. The largest such radical will be denoted by $N$ (rather than a general $U$) and is the unipotent radical of the (minimal parabolic) Borel subgroup $B\subset G$. This will be the main case of interest to us, however, we begin with developing some of the theory for arbitrary unipotent radical $U$ associated with a standard\footnote{A  parabolic subgroup $P\subset G$ is said to be a \emph{standard parabolic} if it contains the Borel subgroup $B$.} parabolic subgroup $P\subset G$, where $P=LU=UL$ is the Levi decomposition and $L$ denotes the Levi factor of $P$.

\subsection{Fourier coefficients}

An important object of interest is the Fourier coefficient $F_{\psi_U}$ associated with a Fourier kernel given by the character (group homomorphism)
\begin{align}
\label{Uchar}
\psi_U\,:\,U(\mathbb{Q})\backslash U(\mathbb{A})\rightarrow U(1)\,.
\end{align}
The notation for the domain indicates that the character is trivial on the discrete subgroup $U(\mathbb{Q})$ in the adelic unipotent $U(\mathbb{A})$ and the image is the circle group of uni-modular complex numbers. The Fourier coefficient $F_{\psi_U}$ of an Eisenstein series $E(\chi,g)$ along $U$ is then defined by the following integral:
\begin{align}\label{FIntegral}
F_{\psi_U}(\chi,g)=\int\limits_{U(\mathbb{Q})\backslash U(\mathbb{A})}E(\chi,ug)\overline{\psi_U(u)}\mathrm{d}u\,.
\end{align}
We denote this more general Fourier coefficient for arbitrary unipotent $U$ by $F_{\psi_U}$ and reserve the notation $W_{\psi}$ for Fourier coefficients/Whittaker vectors (\ref{Wdefintro}) that are defined on the maximal unipotent $N$. 

In general, the unipotent group $U$ can be non-abelian and therefore the character is trivial on the commutator subgroup $U':=[U,U]$. Hence the character can be thought of as defined on the ``abelianisation''  $U'\backslash U$.\footnote{The Lie algebra of (the dual of) this space is called the \textit{character variety}.\label{fn:cv}} For this reason, the Fourier coefficient (\ref{FIntegral}) is sometimes referred to as an abelian Fourier coefficient. It only captures part of the Eisenstein series in the sense that
\begin{align}
\label{abFE}
\sum_{\psi_U} F_{\psi_U} (\chi,g)  = \int\limits_{U'(\mathbb{Q}) \backslash U'(\mathbb{A})} E(\chi,ug) \mathrm{d}u,
\end{align}
where the sum is over all possible characters $\psi_U$ of the type (\ref{Uchar}).\footnote{Interpreted in terms of instanton charges this means that we allow for all mutually local charges, including the zero charge sector of the perturbative theory.} In other words, the Fourier expansion with respect to characters $\psi_U$ does not reflect the dependence of $E(\chi,g)$ on $U'(\mathbb{A})$ as this is averaged out in (\ref{abFE}).

By writing a group element in the form $g =ulk$ with $u\in U$, $l\in L$ and $k\in K$ one finds that
\begin{align}
F_{\psi_U}(\chi,g)=\psi_U(u) F_{\psi_U}(\chi,l)
\label{WhittakerRelation}
\end{align}
and hence $F_{\psi_U}$ is completely determined by its values on the Levi subgroup $L\subset G$. In the following, we will restrict our analysis to this dependence. A function satisfying (\ref{WhittakerRelation}) is known as a {\it (generalized) Whittaker function}.\footnote{The addition ``generalized'' pertains to the fact that the character $\psi_U$ is defined on a unipotent radical which is smaller than the unipotent radical $N$ associated with the minimal (Borel) parabolic $B=NA\subset G$  (see for instance \cite{2012arXiv1210.4064G}). Hence, later when we restrict to characters on $N$ we shall drop the subscript and simply write $\psi$ for $\psi_N$.}

A particular role is played by the trivial character, $\psi_U\equiv 1_U$, given by the identity on $U$.  The corresponding Fourier coefficient represents the zeroth mode of the Fourier expansion, which we denote by $C_U(\chi,g)$. This contribution to the expansion is also referred to as the~\textit{constant term} or \textit{conical vector}~\cite{MR597811}. Each non-trivial character, $\psi_U\neq1_U$, contributes a term $F_{\psi_U}$, in total making up the so-called abelian Fourier coefficients in the expansion of the series. Then the Fourier expansion takes the general form
\begin{align}
E(\chi,g)=C_U(\chi,g)+ \sum_{\psi_U\neq1}F_{\psi_U}(\chi,g)+...\,,
\end{align}
generalising (\ref{Fourierstructure}) from the introduction.
Here, the ellipsis indicates further possible terms associated with the non-zero commutator components of $U$ that are averaged out in (\ref{abFE}). To describe them one has to study non-abelian Fourier expansions, or Fourier-Jacobi expansions, (see, e.g., \cite{VT,Ishikawa,Pioline:2009qt,Bao:2009fg,Persson:2010ms,PerssonAuto,FGKP}) that are associated with the derived series of $U$. In the present article we will, however, not deal with this part of the expansion. As mentioned already, the constant term in the Fourier expansion can be evaluated using Langlands' formula~(\ref{LCF}) or similar formulas derived by M\oe glin--Waldspurger~\cite{Moeglin}. 

The Fourier coefficients $F_{\psi_U}$ possess the important property that their values along $L(\mathbb{Z})$ orbits are related by a simple formula (see e.g.~\cite{GreenSmallRep,MillerSahi}):
\begin{align}
F_{\gamma\cdot \psi_U}(\chi,g) = F_{\psi_U} (\chi,\gamma g) \quad\textrm{for $\gamma\in L(\mathbb{Z})$.}
\end{align}
Here, the action of an element $\gamma$ of the Levi subgroup $L(\mathbb{Z})$ on a character $\psi_U$ is defined by $(\gamma\cdot\psi_U)(u) = \psi_U(\gamma u \gamma^{-1})$. Realising the character in terms of (the dual of) a Lie algebra element of $U'\backslash U$ one is therefore led to the study of character variety orbits in the terminology of footnote~\ref{fn:cv}. These orbits have been completely classified for finite-dimensional simple and simply-laced \textit{complex} Lie algebras~\cite{KostantRallis,Vinberg1975,BalaCarterI,BalaCarterII,Vinberg1979,Littelmann,deGraaf,MillerSahi}; the finer classification for integral rather than complex orbits has only been carried out in some special cases, see for example~\cite{SavinWoodbury,BhargavaI}.

\subsection{Whittaker vectors and characters on $N$}\label{whittchar}

The notion of a Fourier coefficient is general and is used for the $F_{\psi_U}$ making up the abelian part of the Fourier expansion with respect to a unipotent subgroup $U$. From now on we will focus on the case of the so-called minimal parabolic expansion, where $P=B=NA$, such that the unipotent radical is given by $N$. Therefore characters are now group homomorphisms
\begin{align}
\label{charN}
\psi\,:\,N(\mathbb{Q})\backslash N(\mathbb{A})\rightarrow U(1)\,.
\end{align}
Without a subscript characters will always refer to the unipotent $N$ in this article. To further mark the distinction we now use the letter $W_\psi$ and in accordance with standard terminology the Fourier coefficients are defined by the Whittaker integral 
\begin{align}
\label{WIntegral}
W_\psi(\chi,a)=\int\limits_{N(\mathbb{Q})\backslash N(\mathbb{A})}E(\chi,na)\overline{\psi(n)}\mathrm{d}n\
\end{align}
This definition is completely analogous to (\ref{FIntegral}) and we have already restricted the dependence on $G$ to the Levi factor $A$ of the minimal parabolic $B$. The Levi factor in this case is identical to the maximal (split) torus.

It will be important to describe and distinguish in more detail the characters (\ref{charN}) on $N$. To this end we denote by $N_\alpha(\mathbb{A})$ the restriction of the unipotent group $N(\mathbb{A})$ to the one-parameter subgroup associated with the positive root $\alpha$, then we can parametrise the space on which characters $\psi$ depend as
\begin{align}
[N,N]\backslash N\cong \prod_{\alpha\in\Pi} N_\alpha\,.
\end{align}
We denote by $\Pi$ a chosen fixed set of simple roots of $G$.
The character $\psi$ is only sensitive to the part of $N$ in the ``directions'' of the simple roots and we choose to write the character in the following way:
\begin{align}
\label{charM}
\psi\left(\prod_{\alpha\in\Pi}x_\alpha(u_\alpha)\right)=e^{2\pi i\left(\sum_{\alpha\in\Pi}m_\alpha u_\alpha\right)}\,,
\end{align}
where $m_\alpha\in\mathbb{Q}$ are $\mathrm{rk(\mathfrak{g})}$ many parameters that define the character completely.  In the argument of $\psi$ we have used the Chevalley notation 
\begin{align}
\label{chevnot}
x_a(u_\alpha)= \exp(u_\alpha E_\alpha),
\end{align}
 where $E_\alpha$ is the (canonically normalised) step operator corresponding to the (one-dimensional) root space of the simple root $\alpha$. The order of the factors does not matter since $\psi$ is a homomorphism to an abelian group. The $m_\alpha$ parametrise the character variety of footnote~\ref{fn:cv} in this case.

Different values of parameters $m_\alpha$ correspond to different types of the character $\psi$. We distinguish the following three basic types. 
\begin{itemize}
\item[$(i)$] The character is~\textit{trivial} if $m_\alpha=0$ for all $\alpha\in \Pi$ and then $\psi\equiv 1_N$.  
\item[$(ii)$] If $m_\alpha\neq0$ for all $\alpha\in\Pi$, we call the character~\textit{generic} and 
\item[$(iii)$] if $m_\alpha=0$ for at least one, but not all $\alpha\in\Pi$, then the character is~\textit{degenerate}. 
\end{itemize}
We will later use a subset of simple roots $\Pi'\subset\Pi$ to define the non-trivial directions of $\psi$, such that $m_\alpha\neq0$ if $\alpha\in\Pi'$ and is zero otherwise. The character is then said to have support on $\Pi'$. Note that in the following, we also sometimes use the term~\textit{non-generic}, to refer to degenerate characters.

When $\psi$ is degenerate we call the associated integral $W_\psi$ in (\ref{WIntegral}) a \emph{degenerate Whittaker vector}. They represent the main focus of our work. We will deal with this case primarily in section~\ref{degfour}.

\subsection{The Fourier integral}

Let us re-write the expression for the Fourier integral~\eqref{WIntegral} of an expansion with respect to the minimal parabolic subgroup $P=B=NA$. In all manipulations that follow we assume that we are in the region of absolute convergence of sum (\ref{EisA}) defining the Eisenstein series. The first step is to simply substitute the definition of the Eisenstein series (\ref{EisA})
\begin{align}
W_\psi(\chi,a)&=\sum\limits_{\gamma\in B(\mathbb{Q})\backslash G(\mathbb{Q})}\;\int\limits_{N(\mathbb{Q})\backslash N(\mathbb{A})}\chi(\gamma na)\overline{\psi(n)}\mathrm{d}n\,.
\end{align}

We can rewrite the right-hand side as follows (see, e.g., \cite{LanglandsEisenstein,Shahidi})
\begin{align}
\label{step1}
W_\psi(\chi,a)&=\sum\limits_{\gamma\in B(\mathbb{Q})\backslash G(\mathbb{Q})}\;\int\limits_{N(\mathbb{Q})\backslash N(\mathbb{A})}\chi(\gamma na)\overline{\psi(n)}\mathrm{d}n\nn\\
&=\sum\limits_{\gamma\in B(\mathbb{Q})\backslash G(\mathbb{Q})/B(\mathbb{Q})}\; \sum\limits_{\delta\in\gamma^{-1}B(\mathbb{Q})\gamma\cap N(\mathbb{Q})\backslash N(\mathbb{Q})}\;\int\limits_{N(\mathbb{Q})\backslash N(\mathbb{A})}\chi(\gamma \delta na)\overline{\psi(n)}\mathrm{d}n\nn\\
&=\sum\limits_{\gamma\in B(\mathbb{Q})\backslash G(\mathbb{Q})/B(\mathbb{Q})}\; \int\limits_{\gamma^{-1}B(\mathbb{Q})\gamma\cap N(\mathbb{Q})\backslash N(\mathbb{A})}\chi(\gamma na)\overline{\psi(n)}\mathrm{d}n\,.
\end{align}
In the first line, we have written the sum over $\gamma$ in terms of cosets over $B(\mathbb{Q})$ on the right which are labelled by $\delta$. Because of the quotient by $B(\mathbb{Q})$ on the left in the original $\gamma$ sum, we must make sure that we do not overcount the coset representatives $\delta$ and this achieved by the restriction on the $\delta$ sum. In the last step, we have unfolded the sum over $\delta$ to the integration domain by enlarging it. The measure on this larger space is induced from the embedding $N(\mathbb{Q})\to N(\mathbb{A})$. 

As the next step we then use the Bruhat decomposition~(see, e.g., \cite{HumphreysAlgGroup})
\begin{align}
G(\mathbb{Q}) = \bigcup_{w\in \mathcal{W}} B(\mathbb{Q})w B(\mathbb{Q})
\end{align}
to label the double cosets in the $\gamma$ sum in terms of elements of the Weyl group $\mathcal{W}$. 
Then we can rewrite (\ref{step1}) as
\begin{align}\label{fullWhittaker}
W_\psi(\chi,a)=\sum_{w\in\mathcal{W}}\;\int\limits_{w^{-1}B(\mathbb{Q})w\cap N(\mathbb{Q})\backslash N(\mathbb{A})}\chi(wna)\overline{\psi(n)}\mathrm{d}n=\sum_{w\in\mathcal{W}}F_{w,\psi}(\chi,a)\,,
\end{align}
where we defined the short-hand for the individual terms
\begin{align}
\label{FwInt}
F_{w,\psi}(\chi,a)=\int\limits_{w^{-1}B(\mathbb{Q})w\cap N(\mathbb{Q})\backslash N(\mathbb{A})}\chi(wna)\overline{\psi(n)}\mathrm{d}n\,.
\end{align}

\subsection{Integration range}

We begin by analysing in more detail the integration range of the integral in equation~\eqref{FwInt}. Depending on the character $\psi$, we will find that $F_{w,\psi}$ will only be non-zero for a restricted subset of Weyl words.

The integration range of the Fourier integral~\eqref{FwInt} for $F_{w,\psi}$ is given by the coset 
\begin{align}
w^{-1}B(\mathbb{Q})w\cap N(\mathbb{Q})\backslash N(\mathbb{A}).
\end{align}
The intersection in the denominator of this coset consists of those upper elements (generated by positive root generators) of the minimal parabolic subgroup $B$, that are also mapped to upper elements  under the Weyl group action. For the whole denominator we can therefore write
\begin{align}
w^{-1}B(\mathbb{Q})w\cap N(\mathbb{Q})\simeq \prod\limits_{\beta>0|w\beta>0}N_\beta(\mathbb{Q})\,.
\end{align}
With this, the integration range then splits up in the following way
\begin{align}
w^{-1}B(\mathbb{Q})w\cap N(\mathbb{Q})\backslash N(\mathbb{A})\simeq\left(\prod\limits_{\beta>0|w\beta>0}N_\beta(\mathbb{Q})\backslash N_\beta(\mathbb{A})\right)\cdot\left(\prod\limits_{\gamma>0|w\gamma<0} N_\gamma(\mathbb{A})\right)\,.
\end{align}
Let us introduce the following notation. We denote the product in the first parenthesis as
\begin{align}\label{NBeta}
N^w_{\{\beta\}}:=\left(\prod\limits_{\beta>0|w\beta>0}N_\beta(\mathbb{Q})\backslash N_\beta(\mathbb{A})\right)
\end{align}
and the product in the second parenthesis as
\begin{align}\label{NGamma}
N^w_{\{\gamma\}}:=\left(\prod\limits_{\gamma>0|w\gamma<0} N_\gamma(\mathbb{A})\right)\,.
\end{align}
Here the root sets $\{\beta\}$ and $\{\gamma\}$ contain precisely those roots which satisfy the conditions imposed on the products in~\eqref{NBeta} and~\eqref{NGamma}, respectively. A detailed explanation for the particular parametrisation of the set $\{\gamma\}$ is provided in appendix~\ref{param}.

Now, writing for the integration variable $n=n_\beta n_\gamma$ in accordance with this splitting of the integration range a contribution $F_{w,\psi}$ in (\ref{fullWhittaker}) then takes the following form:
\begin{align}
F_{w,\psi}(\chi,a)
&=\int\limits_{N^w_{\{\beta\}}}\int\limits_{N^w_{\{\gamma\}}}\chi(wn_\beta n_\gamma a )\overline{\psi(n_\beta n_\gamma)}\mathrm{d}n_\beta\mathrm{d}n_\gamma\,.
\end{align}
The two integrals can be disentangled further by inserting $w^{-1}w$ between $n_\beta$ and $n_\gamma$ and splitting the Fourier kernel into two factors. One obtains
\begin{align}
F_{w,\psi}(\chi,a)=\int\limits_{N^w_{\{\beta\}}}\int\limits_{N^w_{\{\gamma\}}}\chi(wn_\beta w^{-1}w n_\gamma a)\overline{\psi(n_\beta)}\,\overline{\psi( n_\gamma)}\mathrm{d}n_\beta\mathrm{d}n_\gamma\,.
\end{align}
As the character $\chi$ is left-invariant under any element of $N$ and $wn_\beta w^{-1}\in N$ by the definition of the roots $\beta$ in (\ref{NBeta}) we find
\begin{align}\label{FwSplit}
F_{w,\psi}(\chi,a)=\int\limits_{N^w_{\{\beta\}}}\overline{\psi(n_\beta)}\mathrm{d}n_\beta
\cdot\int\limits_{N^w_{\{\gamma\}}}\chi(w n_\gamma a )\overline{\psi(n_\gamma)}\mathrm{d}n_\gamma\,.
\end{align}
We reiterate from (\ref{NBeta}) and (\ref{NGamma}) that the integration domain $N^w_{\{\beta\}}$ is a compact quotient whereas $N^w_{\{\gamma\}}$ consists of non-compact copies of $\mathbb{A}$ (as many as there are roots $\gamma>0$ with $w\gamma<0$).

\subsection{Conditions for non-zero $F_{w,\psi}(\chi,a)$}\label{nonzero}

The expression~\eqref{FwSplit} gives a restriction on the Weyl words $w$ that yield a non-zero $F_{w,\psi}(\chi,a)$ for a given $\psi$. The reason is that the integral over $n_\beta$ is effectively the average of a character over a full period. If the set $\{\beta\}$ contains one (simple) root along which the character $\psi$ is non-trivial $(m_\beta\neq 0)$ then the character averages to zero. If not, the integral over $N_{\{\beta\}}^w$ yields one by normalisation of the measure on the compact quotient.

Let $\psi$ be a character which is non-trivial along the subset $\Pi'\subset\Pi$ of the set of simple roots $\Pi$, i.e. $m_\alpha\neq 0$ if and only if $\alpha\in \Pi'\subset \Pi$. Considering~\eqref{FwSplit} it is then clear that 
only those Weyl words $w$  that satisfy the condition
\begin{align}
\label{psiwcond}
w\alpha'<0\text{ for all simple roots }\alpha'\in\Pi'
\end{align}
can yield a non-vanishing $F_{w,\psi}$, since otherwise the integral over $N_{\{\beta\}}^w$ vanishes.
We therefore write
\begin{align}\label{whittsum}
W_\psi(\chi,a)=\sum\limits_{w\in\mathcal{W}|w\Pi'<0}F_{w,\psi}(\chi,a)\,,
\end{align}
where in this case
\begin{align}
\label{straightF}
F_{w,\psi}(\chi,a) = \int\limits_{N^w_{\{\gamma\}}} \chi(wn_\gamma a) \overline{\psi(n_\gamma)} \mathrm{d}n_\gamma\,,
\end{align}
since the integral over $N^w_{\{\beta\}}$ in (\ref{FwSplit}) yields unity. \\

An interesting special case is when   $\psi$ is generic since then $\Pi'=\Pi$ and the character $\psi$ is non-trivial along the directions of all simple roots. It is clear that the integral over $n_\beta$ in~\eqref{FwSplit} is then zero (and hence also $F_{w,\psi}$ will be zero), unless the set $\{\beta\}$ does not contain any simple roots. In other words, by the definition~\eqref{psiwcond} all simple roots have to be mapped to negative (simple) roots under the action of $w$. Since all roots are linear combinations of simple roots, this also means that the entire set of positive roots is mapped to negative roots by $w$. 
For finite-dimensional $G$, it is a standard result that this only happens when $w$ is the longest Weyl word, $\wlong$, of the Weyl group of $G$. This means that the sum~\eqref{whittsum} has only one, generically non-zero, contribution coming from $w=\wlong$, and we have
\begin{align}
\label{GenWh}
W_\psi(\chi,a)&=
 \int\limits_{N(\mathbb{A})}\chi(\wlong na )\overline{\psi(n)}\mathrm{d}n.
\end{align}
Here, we have used the fact that $N^{\wlong}_{\{\gamma\}}=N(\mathbb{A})$ and suppressed the index $\gamma$ on $n$ in order to match the standard definition of the generic Whittaker vector in the literature. It is for this (Jacquet--)Whittaker vector that nice simple formulas exist (at the finite places) by the formula of Shintani~\cite{Shintani} or Casselman--Shalika~\cite{CasselmanShalika} (see also \cite{FGKP} for a detailed survey).

\subsection{Whittaker vectors and Kac--Moody groups}
\label{sec_WhittakerKM}

An important observation is that in the case of infinite-dimensional Kac--Moody groups, the expression (\ref{GenWh}) \textit{never} applies. The reason is that there is no longest Weyl word $\wlong$ or no other word that maps all positive simple roots to negative roots. As a result, $F_{w,\psi}$ will be zero whenever the Fourier kernel $\psi$ is generic. The only non-vanishing Whittaker vectors for Kac--Moody groups are therefore those associated with degenerate characters $\psi$. For this reason, and because many interesting special Eisenstein series for finite-dimensional groups are determined by their degenerate Whittaker vectors~\cite{MillerSahi}, we now turn to a more detailed study of degenerate Whittaker vectors.

\section{Degenerate Whittaker vectors}
\label{degfour}

In this section we will present a method for calculating Whittaker vectors $W_\psi(\chi,a)$, when the Fourier kernel $\psi$ is degenerate, cf. section~\ref{whittchar}. For this we will discuss the general schematics of the integral for $F_{w,\psi}$ of (\ref{straightF}), which allows us to derive a reduction formula that expresses $W_\psi$ in terms of non-degenerate Whittaker vectors of subgroups $G'\subset G$ determined by the degenerate character $\psi$.

\subsection{Character twist}
\label{sec_twist}

First we determine the dependence $F_{w,\psi}(\chi,a)$ on $a$. The starting expression is (\ref{straightF}):
\begin{align}
F_{w,\psi}(\chi,a) = \int\limits_{N^w_{\{\gamma\}}} \chi(wn_\gamma a) \overline{\psi(n_\gamma)} \mathrm{d}n_\gamma.
\end{align}
Inserting a factor of $aa^{-1}$ between $w$ and $n_\gamma$ in the argument of $\chi$ and performing a change of integration variables $n\rightarrow a^{-1}na$, under which the measure transforms as $\mathrm{d}n_\gamma\rightarrow\delta_w(a)\mathrm{d}n_\gamma$, we obtain
\begin{align}\label{Fadependence}
F_{w,\psi}(\chi,a)=\chi(waw^{-1})\delta_w(a)\int\limits_{N^w_{\{\gamma\}}}\chi(wn_\gamma)\overline{\psi^a(n_\gamma)}\mathrm{d}n_\gamma\,,
\end{align}
where we have defined the ``twisted'' Fourier kernel $\psi^a(n)=\psi(ana^{-1})$ and we have furthermore extracted the $a$ dependence from the argument of $\chi$. The subscript on the Jacobi factor $\delta_w(a)$ serves to indicate that the integration is not necessarily over all of $N(\mathbb{A})$ and therefore $\delta_w(a)$ is not equal to the standard modulus character $\delta(a)$ of $N(\mathbb{A})$.
As just argued, a non-vanishing $F_{w,\psi}$ is expressed solely in terms of an integral over $N^w_{\{\gamma\}}$ and it is this part of all of $N(\mathbb{A})$ that contributes to $\delta_w(a)$. That is, $\delta_w(a)$ is given by the relation $\mathrm{d} (an_\gamma a^{-1}) =\delta_w(a) \mathrm{d}n_\gamma$, where $n_\gamma$ is an element of $N^w_{\{\gamma\}}$ as before. With (\ref{NGamma}), one has
\begin{align}
\label{deltaw}
\delta_w(a) = a^{\sum_{\gamma>0\,|\, w\gamma<0} \gamma} = a^{\rho-w^{-1}\rho},
\end{align}
where we have used standard results on the set $\{\gamma>0\,|\,w\gamma<0\}$ given for example in~\cite{Liu}, that we also rederive in appendix~\ref{param} for completeness. Using also the expression~\eqref{chilambda} for $\chi(waw^{-1})$, we deduce that the prefactor in \eqref{Fadependence} is given by
\begin{align}\label{prefactor}
\chi(waw^{-1})\delta_w(a) = a^{w^{-1}(\lambda+\rho)+\rho-w^{-1}\rho} = a^{w^{-1}\lambda + \rho}.
\end{align}
In order to ease the notation in the following chapters, let us define the integral
\begin{align}\label{Fw}
\mathcal{F}_{w,\psi}(\chi):=F_{w,\psi}(\chi, \id)=\int\limits_{N^w_{\{\gamma\}}}\chi(wn_\gamma)\overline{\psi(n_\gamma)}\mathrm{d}n_\gamma\,.
\end{align}
In terms of this quantity the full Whittaker vector (\ref{fullWhittaker}) is then
\begin{align}
\label{WpsiF}
W_\psi(\chi,a) = \sum_{w\in\mathcal{W}} a^{w^{-1}\lambda+\rho} \mathcal{F}_{w,\psi^a}(\chi),
\end{align}
where the twisted character $\psi^a$ enters. Below we will study in detail the integral (\ref{Fw}) for $\mathcal{F}_{w,\psi}$ for arbitrary $\psi$ and only substitute back the particular twisted character $\psi^a$ at the very end of the calculation.

\subsection{Parametrising the contributing Weyl words}
\label{sec_WeylWords}

According to the discussion in section~\ref{nonzero}, the contributing set of Weyl words for which $\mathcal{F}_{w,\psi}$ can be non-zero is given by
\begin{align}\label{specialw}
\mathcal{C}_\psi:=\{w\in\mathcal{W}\,|\,w\Pi'<0\}\,,
\end{align}
where $\Pi'\subset\Pi$ denotes the simple roots $\alpha$ for which $m_\alpha\neq 0$ (cf. also~(\ref{charM})). We have added the subscript $\psi$ as a reminder that the definition of the set depends on $\psi$.

We are now going to characterise the special set $\mathcal{C}_\psi$ of Weyl words~\eqref{specialw} in a more practical way. The set of simple roots $\Pi'$, together with its complement $\overline{\Pi'}$, partition the set of simple roots $\Pi$ of $G$. The subgroup $G'$ of the full invariance group $G$ is defined by the Dynkin diagram given by $\Pi'$ and we assume it to be finite-dimensional. The Weyl group associated with $\Pi'$ is denoted by $\mathcal{W}'$.

The statement that we are going to prove in the following is that the elements of our special set of words can be written in the following form\footnote{We recall that a given Weyl element $w\in\mathcal{W}$ has of course many seemingly different representations in terms of products of other elements. What we are claiming here is that all $w$ that satisfy $w\Pi'<0$ have a representation in the form given.}
\begin{align}\label{wordsplit}
w \in \mathcal{C}_\psi\quad\Longleftrightarrow \quad w=w_c\wlong'\,,
\end{align}
where $\wlong'$ is the longest Weyl word of $\mathcal{W}'$ and $w_c$ is a carefully chosen representative of the coset $\mathcal{W}/\mathcal{W}'$. We refer to the construction of the particular coset representative that we require as the~\textit{orbit method} and we will outline it in section~\ref{orbit}. It is analogous to the method explained in~\cite{FK2012}. Before we present this construction, let us first characterise the representative $w_c$ required in (\ref{wordsplit}).

By its very definition, the action of the longest Weyl word $\wlong'$ of $\mathcal{W}'$ makes all roots of $G'$, and in particular the simple roots in $\Pi'$, negative. In order to satisfy the condition (\ref{specialw}), one can then add further Weyl words $w_c$ to the left of $\wlong'$, provided they satisfy the condition
\begin{align}
\label{wccond}
w_c\alpha'>0\text{ for all }\alpha'\in \Pi'\,.
\end{align}
Then the combined word $w=w_c \wlong'$ will map all simple roots in $\Pi'$ to negative roots. Weyl words $w_c$ satisfying (\ref{wccond}) can be constructed as carefully chosen representatives of the coset $\mathcal{W}/\mathcal{W}'$. We will say more about the construction of these Weyl words in section~\ref{orbit}.

\subsection{Reduction formula}

We now return to evaluating the contribution $\mathcal{F}_{w,\psi}$  to a degenerate Whittaker vector given by (\ref{FwSplit}) and use that $w=w_c\wlong'\in\mathcal{C}_\psi$ with the particular parametrisation of the preceding section. 
Associated with the word $w=w_c\wlong'$ is a parametrisation of the elements $n_\gamma$ in the integration domain $N^w_{\{\gamma\}}$, cf.~(\ref{NGamma}). We write 
\begin{align}
\label{ndecomp}
n_\gamma = n_c n' \quad\textrm{with}\quad  n_c\in N_c(\mathbb{A}) \textrm{ and } n'\in N'(\mathbb{A}),
\end{align}
where $N'(\mathbb{A})$ is the unipotent subgroup of the minimal Borel $B'(\mathbb{A})$ of the subgroup $G'(\mathbb{A})$ determined by the set of simple roots $\Pi'$ that indicate the directions on which the degenerate character $\psi$ depends non-trivially. The set $N_c(\mathbb{A})$ involves the remaining positive roots $\gamma$ that are mapped to negative roots by the action of $w$ but that are not positive roots of the subgroup $G'$.  The degenerate character $\psi$ does not depend on $n_c$ since $N_c(\mathbb{A})$ does not involve any of the simple roots of $G'$. 
We can thus write the integral for $\mathcal{F}_{w,\psi}$ as
\begin{align}
\mathcal{F}_{w,\psi}(\chi)=\int\limits_{N_c(\mathbb{A})}\int\limits_{N'(\mathbb{A})} \chi(w_c \wlong' n_c n') \overline{\psi(n')} \mathrm{d}n_c \mathrm{d}n' \,.
\end{align}
The argument of the character $\chi$ can be rewritten as
\begin{align}
\chi(w n_c n') = \chi (wn_cw^{-1} w n') 
= \chi(wn_cw^{-1} w_c \tilde{n}\tilde{a}),
\end{align}
where we have used $w=w_c\wlong'$. Moreover, $\tilde{n}\tilde{a}\tilde{k}=\wlong' n'$ arises from the Iwasawa decomposition in the group $G'$ and we have used the right-invariance of $\chi$ under $\tilde{k}$ in the last step. Now, it is important that $w_c$ satisfies the condition (\ref{wccond}) which implies that, even though it is no longer in $G'(\mathbb{A})$, the element $w_c\tilde{n}\tilde{a}w_c^{-1}=\hat{n}\hat{a}$ is an element of the Borel subgroup $B(\mathbb{A})$ with $\hat{n}\in N(\mathbb{A})$ and $\hat{a}\in A(\mathbb{A})$. Hence, in particular $\hat{a}=w_c\tilde{a}w_c^{-1}$. In the next step we conjugate both $\hat{n}$ and $\hat{a}$ through to the left in the argument of $\chi$. For $\hat{n}$ this induces a uni-modular change of integration variables for $n_c$~\cite{Liu}. By contrast, for $\hat{a}$ this generates a non-trivial Jacobi factor when passing past $wn_cw^{-1}$ that we can determine in a way similar to (\ref{deltaw}). The relevant manipulation is:
\begin{align}
\int\limits_{N_c(\mathbb{A})} \chi(\hat{n}wn_cw^{-1} \hat{a}) \mathrm{d}n_c 
&= \int\limits_{N_c(\mathbb{A})} \chi(\hat{n}\hat{a}wn_cw^{-1} ) \hat{a}^{w_c\rho-\rho} \mathrm{d}n_c \nn\\
&= \int\limits_{N_c(\mathbb{A})} \chi(wn_cw^{-1} ) \tilde{a}^{w_c^{-1}\lambda+\rho} \mathrm{d}n_c ,
\end{align}
where we have used $\chi(\hat{n}\hat{a})=\hat{a}^{\lambda+\rho}$ and $\hat{a}=w_c\tilde{a}w_c^{-1}$. Now, $\tilde{a}$ does not depend on $n_c$ and we can take it out of the $N_c(\mathbb{A})$ integral. None of these transformations have any impact on the argument of the character $\psi(n')$.

The result of these steps is that $\mathcal{F}_{w,\psi}$ factorises according to
\begin{align}\label{intfactor}
\mathcal{F}_{w,\psi}(\chi)= \int\limits_{N_c(\mathbb{A})} \chi(w_c\wlong'n_c) \mathrm{d}n_c \cdot \int\limits_{N'(\mathbb{A})} \chi'(\wlong' n') \overline{\psi(n')} \mathrm{d}n',
\end{align}
where the character $\chi':B'(\mathbb{A})\to \mathbb{C}^*$ is given by the (projection of the) weight $w_c^{-1}\lambda+\rho$. A different proof for this factorisation is given in appendix~\ref{sec_proof}. The Jacobi factor arising from $\hat{a}$ has been transformed back into the expression $\chi'(\wlong'n')$ in the second integral.

The two separate integrals in~\eqref{intfactor} are of well-known types. The $N_c(\mathbb{A})$ integral is identical to the integral that determines (the numerical coefficient of) the contribution of the Weyl word $w_c$ to the constant term  (in the minimal parabolic)  (\ref{Mwint}) and therefore yields the factor $M(w_c^{-1},\lambda)$. From~\eqref{ndecomp}, we recall that $\wlong' N_c(\mathbb{A}) (\wlong')^{-1}$ is the product of one-parameter subgroups whose roots are mapped to negative roots by $w_c$.  Referring back to (\ref{GenWh}), we recognise the second integral in~\eqref{intfactor} as the non-degenerate Whittaker vector for the subgroup $G'(\mathbb{A})\subset G(\mathbb{A})$ for a \textit{generic} Fourier character $\psi$ on $N'$ of the Eisenstein series determined by the weight $w_c^{-1}\lambda+\rho$, projected orthogonally to $G'(\mathbb{A})$, and evaluated at the identity.
The expression~\eqref{intfactor} for $\mathcal{F}_{w,\psi}$, then evaluates to
\begin{align}
\label{finalF}
\mathcal{F}_{w,\psi}(\lambda)=M(w_c^{-1},\lambda)W'_{\psi}(w_c^{-1}\lambda,\id)\,.
\end{align}

Equation (\ref{finalF}) is the expression for an arbitrary character $\psi$. For the Whittaker vector $W_\psi$ we require (\ref{finalF}) evaluated at the twisted character $\psi^a$ according to (\ref{WpsiF}). Combining all elements as prescribed by~\eqref{WpsiF} we obtain the following final expression for the degenerate Whittaker vector, claimed in proposition~\hyperref[prop1]{\ref*{prop1}},
\begin{align}\label{degW1}
W_\psi (\lambda,a) = \sum_{w_c\wlong'\in\, \mathcal{C}_\psi} a^{(w_c\wlong')^{-1}\lambda+\rho} M(w_c^{-1},\lambda) W'_{\psi^a}(w_c^{-1}\lambda,\id)\,,
\end{align}
where the factor in front of $\mathcal{F}_{w,\psi^a}$ was determined in (\ref{prefactor}).
Here, $W'_{\psi^a}$ denotes a Whittaker function of the $G'(\mathbb{A})$ subgroup of $G(\mathbb{A})$. Therefore, Whittaker vectors of degenerate characters $\psi$ can be evaluated as sums over Whittaker vectors of subgroups on which the character is generic.  At non-archimedean places these then can be evaluated using the Shintani--Casselman--Shalika formula~\cite{Shintani,CasselmanShalika}.

\section{Whittaker vectors for Kac--Moody groups}
\label{sec_KMWhittaker}

In this section we discuss the calculation of Whittaker vectors of Kac--Moody Eisenstein series. Due to the arguments of section~\ref{sec_WhittakerKM}, we will only be concerned with the computation of Whittaker vectors of degenerate type.

Let us outline the central problem in an explicit evaluation of the reduction formula~\eqref{degW1} in the case of Kac--Moody Eisenstein series. Namely the obvious problem is that the set of contributing Weyl words, a~priori appears to be infinite, due to the infinite-dimensional nature of the Kac--Moody group and its associated Weyl group. For the moment we will keep our discussion general and specialise to particular types of Kac--Moody groups later on.

As discussed in section \ref{sec_background}, the same problem also appears in Langlands' formula for the constant term. However, as was shown in~\cite{FK2012}, this problem is resolved when applying the formula for certain special types of Eisenstein series. These series are the maximal parabolic Eisenstein series, defined around (\ref{maxlambda}), with a special choice for the maximal parabolic subgroup and the parameter $s$ defining the series. In these cases the apparently infinite series for the constant term ``collapses'' to a~\textit{finite} (and indeed very small) number of terms. In particular the series for which this happens, are the Eisenstein series, which appear as the automorphic couplings (up to a constant factor) of the lowest orders of curvature corrections in the low-energy expansion of type II string theory (see section \ref{sec_background} for more information and references). As displayed in (\ref{maxlambda}), these series, defined with respect to the maximal parabolic subgroup $P_{i_*}$, have as defining weight $\lambda=2s\Lambda_{i_*}-\rho$ and hence are of the form:
\begin{align}
E(2s\Lambda_{i_*}-\rho,g)\,.
\end{align}
For the choice $i_*=1$, $s=3/2$ and $5/2$, this series appears in the coefficient of the $\mathcal{R}^4$ and $\partial^4\mathcal{R}^4$ curvature correction, respectively. Motivated by this, the collapse of the constant terms of these series was demonstrated in~\cite{FK2012}, for the Kac--Moody groups $E_9$, $E_{10}$ and $E_{11}$, i.e. in $D=2$, $1$ and $0$ dimensions, respectively. (The $E_n$ Dynkin diagram with our labelling conventions of the roots was given in Figure~\ref{fig:dynkin} in the introduction.)

The collapse of the constant term is generally encoded in the factor $M(w,\lambda)$ appearing in (\ref{LCF}) and can be shown to vanish for all but a finite number of Weyl words $w$ for these special choices of $s$ and $i_*$. As the same factor also appears in the reduction formula~\eqref{degW1}, one therefore expects to observe a similar collapse in the degenerate Whittaker vectors. We now summarise the general mechansim for collapse in some detail, by outlining important properties of the $M(w,\lambda)$ coefficient. For full details, we refer the reader to~\cite{FK2012}.

\subsection{The collapse mechanism}

For the reader's convenience, we state again the definition of the $M(w,\lambda)$ factor (\ref{Mw}):
\begin{align}
M(w,\lambda)=\prod_{\alpha>0|w\alpha<0}\frac{\xi(\langle\lambda|\alpha\rangle)}{\xi(1+\langle\lambda|\alpha\rangle)}=\prod_{\alpha>0|w\alpha<0} c\left(\langle\lambda|\alpha\rangle\right)
\end{align}
where we defined $c(k):=\xi(k)/\xi(1+k)$. The function $c(k)$ only has special values at $k=\pm1$, with
\begin{align}
c(-1)=0\quad\text{and}\quad c(+1)=\infty,
\end{align}
corresponding to a simple zero and a simple pole. (This means that products that are formally of the type $c(+1)c(-1)$ have finite limits when $s$ approaches one of its special values since $s$ only appears linearly in the argument of $c(k)$.) From these basic facts we conclude that for a given Weyl word $w\in\mathcal{W}$, the product making up $M(w,\lambda)$ will be zero, if there are~\textit{more} $c(-1)$ factors than $c(+1)$, appearing in it. Furthermore, by the multiplicative property (\ref{Mmult}), we conclude that if $M(\tilde{w},\lambda)=0$ for a given Weyl word $\tilde{w}$ then the factor $M(w\tilde{w},\lambda)=M(w,\tilde{w}\lambda)M(\tilde{w},\lambda)$, associated with any longer Weyl word of the form $w\tilde{w}$, will also be zero~\cite{Green:2010kv} as long as $M(w\tilde{w},\lambda)$ is finite. 

For maximal parabolic series at special values of $s$, the truncation of the apparently infinite sum over Weyl words in Langlands formula and the reduction formula for Whittaker vectors, through the factor $M(w,\lambda)$ occurs in two steps, which we will now discuss.

\subsubsection{Maximal parabolic criterion}

Restricting to maximal parabolic series with defining weight~$\lambda=2s\Lambda_{i_*}-~\rho$, it is easy to see then that $\langle\alpha_i|\lambda\rangle=-1$ for all simple roots $\alpha_i\neq\alpha_{i_*}$. In case a simple root other than $\alpha_{i_*}$ appears in the product for $M(w,\lambda)$, it will introduce a factor $c(-1)=0$ and the product can easily be shown to vanish. Non-vanishing $M(w,\lambda)$ are obtained by restricting to the following set of Weyl words
\begin{align}
\label{SPistar}
\mathcal{S}_{\Pi^*}=\{w\in\mathcal{W}|w\alpha_i>0\text{ for all }\alpha_i\in\Pi^*\}\,,
\end{align}
where $\Pi^*:=\Pi\backslash \{\alpha_{i_*}\}$. This set of Weyl words, can be constructed by the ``orbit method'' which we will describe in section~\ref{orbit}. It represents the set of minimal reflections needed to construct the Weyl orbit of the fundamental weight $\Lambda_{i_*}$.\footnote{Note that in~\cite{FK2012} this set was denoted by $\mathcal{S}_{i_*}$} We can think of $\mathcal{S}_{\Pi^*}$ as being the set of minimal coset representatives of $\mathcal{W}/\mathcal{W}^*$, where $\mathcal{W}^*$ denotes the Weyl subgroup of $\mathcal{W}$ generated by the fundamental reflections in the simple roots $\Pi^*$ only.
Let us point out that the set $\mathcal{S}_{\Pi^*}$ only contains a fraction of the total set of Weyl words $\mathcal{W}$. Nevertheless, it is important to note that for the case of Kac--Moody groups, the set $\mathcal{S}_{\Pi^*}$, still contains an infinite number of elements. 
In the second step, which we will now explain, this infinite number will be further reduced, to a finite,  small number, via special choices of the parameter $s$.

\subsubsection{Special $s$ and $i_*$ values}
\label{sistar}

Restricting to special values for $s$ and $i_*$, we see that there are two special sets of positive roots $\Delta_{s,i_*}(\pm1)$, defined as
\begin{align}
\Delta_{s,i_*}(\pm1):=\{\alpha\in\Delta_+\;:\;\langle\lambda|\alpha\rangle=\langle2s\Lambda_{i_*}-\rho|\alpha\rangle=\pm1\}\,,
\end{align}
where $\Delta_+$ is the set of all positive roots of $G$. We stress that these sets are not necessarily finite for arbitrary choices of $i_*$ and $s$ in the Kac--Moody case. For example, if there is a null root $\delta$ and supposing it satisfies $\langle \lambda|\delta\rangle=0$, then the sets will be trivially of infinite order. This would happen for example for $E_{10}$ and $i_*=10$ for arbitrary values of $s$.

The task now is to check how many of the roots contained in $\Delta_{s,i_*}(\pm 1)$ will contribute to $M(w,\lambda)$ for a given Weyl word $w$. 
On a practical level, this is done by parsing through the Weyl orbit of $\Lambda_{i_*}$ by increasing length of the Weyl word. For each Weyl word $w$ we check the number of $c(-1)$ factors and the number of $c(+1)$ factors that contribute to the product of $M(w,\lambda)$. As discussed above, a Weyl word will only yield a non-zero $M(w,\lambda)$ factor, provided the number of $c(-1)$ factors, is smaller or equal to the number of $c(+1)$ factors. By the multiplicative property of $M(w,\lambda)$, we know that once it vanishes for a certain $\tilde{w}$ in a branch, we need not consider Weyl words $w\tilde{w}$ that ``end on'' $\tilde{w}$.

It turns out that for specific choices of the parameters $s$ and $i_*$, only a finite number of the Weyl words $\mathcal{S}_{\Pi^*}$ contribute, also in the case of Kac--Moody groups. As already mentioned above, some particular values for which this is the case are $i_*=1$ and $s=3/2$ or $5/2$ and these are the ones relevant in string theory. In~\cite{FK2012}, it was shown that for the $E_9$, $E_{10}$ and $E_{11}$ maximal parabolic Kac--Moody Eisenstein series with $i_*=1$, there are other (small) integer and half-integer values of $s$ for which this collapse happens. The results are summarised in table 4 of that publication. 

\subsection{Orbit method}\label{orbit}

In the following we will describe a general method for constructing a set of Weyl words which satisfy the condition:
\begin{align}
w\alpha_i>0\text{ for all simple roots $\alpha_i\in \Pi^*$}\,.
\end{align}
It thus provides a way of computing the sets of Weyl words $\mathcal{S}_{\Pi^*}$ in (\ref{SPistar}), as well as the Weyl words $w_c$ satisfying condition~\eqref{wccond} that enter in the set $\mathcal{C}_\psi$ determined by a (degenerate) character $\psi$. 

The construction proceeds in the following way. Consider the dominant weight
\begin{align}
\Lambda=\sum_{\alpha\in\overline{\Pi^*}}\Lambda_{\alpha}\,.
\end{align}
Its $\mathcal{W}$-orbit points are in bijection with the coset $\mathcal{W}/\mathcal{W}^*$ as $\mathcal{W}^*$ stabilises $\Lambda$.\footnote{For the case (\ref{SPistar}) this means $\Lambda=\Lambda_{i_*}$.} We construct its orbit under the action of the Weyl group $\mathcal{W}$ of $G$ iteratively, according to the following standard algorithm:
\begin{enumerate}
\item Start with the initial set of orbit points $\mathcal{O}=\{\Lambda\}$.
\item Given a weight $\mu\in\mathcal{O}$, compute its Dynkin labels $p_\alpha=\langle \mu |\alpha\rangle$ for all $\alpha\in \Pi$. 
\item For all labels $p_\alpha$ that are strictly positive, construct $\mu'=w_\alpha \mu$ where $w_\alpha$ is the fundamental reflection in the simple root $\alpha$. Add the resulting $\mu'$ to the set $\mathcal{O}$ of orbit points if they are not already in there.
\item If there remains a weight $\mu$ in $\mathcal{O}$ for which steps 2. and 3. have not been carried out, repeat them for this $\mu$.
\end{enumerate}
This algorithm constructs the orbit representatives of the $\mathcal{W}$-orbit of $\Lambda$. If one remembers for each orbit point $\mu$ in the orbit the sequence of fundamental reflections that were needed to obtain it, one thus obtains a set of minimal (with respect to word length) Weyl words that relate the dominant weight $\Lambda$ to each of its images.  (For an illustration of this method, see section 3.3 of~\cite{FK2012}.) 
With this algorithm one can construct exactly the minimal Weyl words that for example make up the set $\mathcal{S}_{\Pi^*}$ or satisfy condition~\eqref{wccond}.

We note that in the case of Kac--Moody groups, the Weyl orbit of $\Lambda$ is of infinite size and the algorithm has to be truncated in practice. However, the algorithm provides the Weyl words of $\mathcal{S}_{\Pi^*}$ as a partially ordered set of longer and longer Weyl words that get longer on the left as one moves away from the dominant $\Lambda$. Due to the multiplicative property (\ref{Mmult}) one does not need to continue the Weyl orbit past a Weyl word whose $M$-factor vanishes. 
For the special Eisenstein series of interest in string theory only a finite number of Weyl words in $\mathcal{S}_{\Pi^*}$ remain.
These finitely many terms can be calculated explicitly. 

We denote the set of Weyl words $w$ that have a non-zero factor $M(w,\lambda)$ by
\begin{align}
\label{Clambda}
\mathcal{C}_\lambda = \left\{ w\in \mathcal{W}\,|\, M(w,\lambda)\neq 0\right\}\subset \mathcal{S}_{\Pi*},
\end{align}
where we also allow potentially infinite values. In the final expressions, these always appear in combinations such that the sum over them has a well-defined limit.

\subsection{The collapse for degenerate Whittaker vectors}

As the next step, we will now combine the collapse mechanism with the formula (\ref{degW1}) for degenerate Whittaker vectors. This will allow us to calculate explicitly some degenerate Whittaker vectors for Kac--Moody groups. The following applies to (maximal) parabolic Eisenstein series.

Looking at the reduction formula (\ref{degW1}), we first construct the set $\mathcal{C}_\lambda$ defined in (\ref{Clambda}). In the context of (\ref{degW1}) the Weyl elements $w$ of $\mathcal{C}_\lambda$ with non-vanishing $M(w,\lambda)\neq 0$ should be interpreted as $w_c^{-1}$. We also know that all words $w\in \mathcal{C}_\psi$ contributing to (\ref{degW1}) are of the form $w=w_c\wlong'$, cf.~(\ref{wordsplit}). Therefore we form the set 
\begin{align}
\mathcal{C}_{\lambda,\psi} = \left\{ w_c^{-1} \in \mathcal{C}_\lambda \,|\, w_c\wlong'\in \mathcal{C}_\psi\right\}
=\mathcal{C}_\psi \cap \left((\mathcal{C}_\lambda)^{-1}\wlong'\right),
\label{collapsingset}
\end{align}
where $(\mathcal{C}_\lambda)^{-1} \wlong'$ denoted the set of the inverses of all $\mathcal{C}_\lambda$ elements and then multiplied on the right by $\wlong'$. The sum in the reduction formula~\eqref{degW1} is restricted to $\mathcal{C}_{\lambda,\psi}$.

The set $\mathcal{C}_{\lambda,\psi}$ typically contains a very small number of elements for the special values of $\lambda$, i.e. $i_*$ and $s$, that are relevant in string theory. We stress that it is not guaranteed that the set $\mathcal{C}_{\lambda,\psi}$ is finite. Only for very special choices of $\lambda$ we expect finite sets and these are the ones that have appeared in string theory thus far. The analysis of~\cite[Table 4]{FK2012} revealed a few additional values of $s$ where simplifications occur. There is no systematic understanding of  ``good'' choices of $i_*$ and $s$ at the moment; we offer a few more comments on this point in the conclusions.

The reduction to $\mathcal{C}_{\lambda,\psi}$ is not the only simplification that arises. Further terms can be absent for a degenerate Whittaker vector when the factor $W_{\psi^a}'((w_c^{-1}\lambda)_{G'},\id)$ vanishes and its prefactor is finite. This vanishing of $W_{\psi^a}'((w_c^{-1}\lambda)_{G'},\id)$ happens for example always when the projected weight $(w_c^{-1}\lambda)_{G'}$ is equal to $-\rho'$ ($\rho'$ being the Weyl vector of $G'$). The reason is that for this case one is computing the Whittaker vector of a constant function which vanishes. Similar cases can arise when $(w_c^{-1}\lambda)_{G'}$ is such that generic Whittaker vector on $G'$ vanishes, i.e., the projected weight corresponds to a degenerate principal series representation. The application of this criterion is a bit more subtle and typically involves the analysis of a family of maximal parabolic Eisenstein series.

\section{Explicit results for some Kac--Moody groups}
\label{sec_KMtables}

We now present the results that we obtained by implementing the formalism of section~\ref{sec_KMWhittaker} for $E_9$, $E_{10}$ and $E_{11}$ for the special cases $i_*=1$ and $s=3/2$ and $s=5/2$ in (\ref{maxlambda}). See figure~\ref{fig:dynkin} for our labelling convention of the $E_n$ Dynkin diagram.

We present the results in table form and use some short-hand notations for the Whittaker vectors of the subgroups $A_1$ and $A_2$ that arise. We denote an element of the maximal torus by
\begin{align}
a= \prod_{i=1}^{n} v_i^{h_{i}},
\end{align}
where $h_{i}$ is the standard Chevalley generator of the simple root $\alpha_i$ in Bourbaki labelling and $n$ is the rank of the group. In the case of the affine Kac--Moody group $E_9$ an element of the maximal torus is of the form $av^d$, where $d$ is the derivation element in the Cartan subalgebra~\cite{GarlLoop}.

\subsection{The case $s=3/2$ or $\mathcal{R}^4$}

The implementation of the reduction formula (\ref{degW1}) shows that the Whittaker vectors of the Eisenstein series with $(i_*,s)=(1,\frac32)$ are only non-zero when $G'=SL(2,\mathbb{A})$. In other words, the character $\psi$ is maximally degenerate (without being trivial).
Therefore the set $\Pi'$ of section~\ref{degfour} contains only one simple root. For such a degenerate character $\psi$ with only non-zero charge $m_\alpha$ for a single simple root $\alpha$ we write the corresponding $A_1$-type Whittaker vector as 
\begin{align}
W'_{\psi^a}(\chi',\id) &:= B_{s',m_\alpha}(a^\alpha) \\
&:= \frac{2}{\xi(2s')}|a_\alpha|^{s'-1/2} |m_\alpha|^{1/2-s'} \sigma_{2s'-1}(m_\alpha)  K_{s'-1/2}(2\pi |m_\alpha| a^\alpha).\nn
\end{align}
Here, $s'$ parametrises the projected character $\chi'$ on $SL(2)$ by $\lambda'=2s'\Lambda'-\rho'$ where $\Lambda'$ is the unique fundamental weight of $SL(2)$. In order to obtain finite coefficients we will also use the differently normalised Bessel-type function 
\begin{align}
\tilde{B}_{s',m_\alpha}:=\xi(2s')B_{s',m_\alpha}
\end{align}
 for some entries of the table.

\subsubsection{$E_9$ with $s=3/2$}

As shown in~\cite{FK2012}, the series that appears in string theory has to be multiplied by an overall factor of $v$. This means that the function that should appear in string theory is 
\begin{align}
f_{E_9}^{(0)}(g) = v E(3\Lambda_1-\rho,g),
\end{align}
where the notation of (\ref{corrections}) in the introduction was used. We will not display the extra $v$ in the table that therefore only contains the Whittaker vectors of $E(3\Lambda_1-\rho,g)$.

Even though the Cartan subalgebra of $E_9$ is ten-dimensional, the characters $\psi$ on $N$ are only labelled by nine parameters $m_\alpha$ since there are only nine simple roots.

\begin{center}
\renewcommand{\arraystretch}{1.6}
\begin{tabular}{ | c || c | }
  \hline                       
 $\psi$ & $W_{\psi}(\chi_{3/2},a)$ \\[-0.07cm]\hline \hline
 $(m,0,0,0,0,0,0,0,0)$ & $v_3^2 v_1^{-1}B_{3/2,m}\left(v_1^2v_3^{-1}\right)$ \\ \hline
 $(0,m,0,0,0,0,0,0,0)$ & $\frac{v_2^2\tilde{B}_{0,m}(v_2^2v_4^{-1})}{\xi(3)}$ \\ \hline
 $(0,0,m,0,0,0,0,0,0)$ & $\frac{\xi (2) v_4 B_{1,m}\left( v_3^2 v_1^{-1}v_4^{-1}\right)}{\xi (3)}$ \\ \hline
 $(0,0,0,m,0,0,0,0,0)$ & $\frac{v_4\tilde{B}_{1/2,m}(v_4^2v_2^{-1}v_3^{-1}v_5^{-1})}{\xi(3)}$  \\ \hline
 $(0,0,0,0,m,0,0,0,0)$ & $\frac{v_5^2\tilde{B}_{0,m}(v_5^2v_4^{-1}v_6^{-1})}{\xi(3)v_6}$ \\ \hline
 $(0,0,0,0,0,m,0,0,0)$ & $\frac{\xi (2) v_6^3 B_{-1/2,m}\left(v_6^2 v_5^{-1}v_7^{-1}\right)}{\xi (3)v_7^2}$  \\ \hline
 $(0,0,0,0,0,0,m,0,0)$ & $ v_7^4v_8^{-3} B_{-1,m}\left(v_7^2 v_6^{-1}v_8^{-1}\right)$  \\ \hline
 $(0,0,0,0,0,0,0,m,0)$ & $\frac{\xi (4) v_8^5v_9^{-4} B_{-3/2,m}\left(v_8^2 v_7^{-1}v_9^{-1}\right)}{\xi (3)}$  \\ \hline
 $(0,0,0,0,0,0,0,0,m)$ & $\frac{\xi (5) v_9^6v^{-5} B_{-2,m}\left(v_9^2 v_8^{-1}v^{-1}\right)}{\xi (3)}$  \\ \hline
 \end{tabular}
\end{center}

\subsubsection{$E_{10}$ with $s=3/2$}

\begin{center}
\renewcommand{\arraystretch}{1.6}
\begin{tabular}{ | c || c | }
  \hline                       
  $\psi$ & $W_{\psi}(\chi_{3/2},a)$ \\[-0.07cm] \hline \hline
 $(m,0,0,0,0,0,0,0,0,0)$ & $v_3^2 v_1^{-1}B_{3/2,m}\left(v_1^2v_3^{-1}\right)$ \\ \hline
 $(0,m,0,0,0,0,0,0,0,0)$ & $\frac{v_2^2\tilde{B}_{0,m}(v_2^2v_4^{-1})}{\xi(3)}$ \\ \hline
 $(0,0,m,0,0,0,0,0,0,0)$ & $\frac{\xi (2) v_4 B_{1,m}\left( v_3^2 v_1^{-1}v_4^{-1}\right)}{\xi (3)}$ \\ \hline
 $(0,0,0,m,0,0,0,0,0,0)$ & $\frac{v_4\tilde{B}_{1/2,m}(v_4^2v_2^{-1}v_3^{-1}v_5^{-1})}{\xi(3)}$  \\ \hline
 $(0,0,0,0,m,0,0,0,0,0)$ & $\frac{v_5^2\tilde{B}_{0,m}(v_5^2v_4^{-1}v_6^{-1})}{\xi(3)v_6}$ \\ \hline
 $(0,0,0,0,0,m,0,0,0,0)$ & $\frac{\xi (2) v_6^3 B_{-1/2,m}\left(v_6^2 v_5^{-1}v_7^{-1}\right)}{\xi (3)v_7^2}$  \\ \hline
 $(0,0,0,0,0,0,m,0,0,0)$ & $ v_7^4v_8^{-3} B_{-1,m}\left(v_7^2 v_6^{-1}v_8^{-1}\right)$  \\ \hline
 $(0,0,0,0,0,0,0,m,0,0)$ & $\frac{\xi (4) v_8^5v_9^{-4} B_{-3/2,m}\left(v_8^2 v_7^{-1}v_9^{-1}\right)}{\xi (3)}$  \\ \hline
 $(0,0,0,0,0,0,0,0,m,0)$ & $\frac{\xi (5) v_9^6v_{10}^{-5} B_{-2,m}\left(v_9^2 v_8^{-1}v_{10}^{-1}\right)}{\xi (3)}$  \\ \hline
 $(0,0,0,0,0,0,0,0,0,m)$ & $\frac{\xi (6) v_{10}^7 B_{-5/2,m}\left(v_{10}^2 v_9^{-1}\right)}{\xi (3)}$  \\ \hline
 \end{tabular}
\end{center}

\subsubsection{$E_{11}$ with $s=3/2$}

\begin{center}
\renewcommand{\arraystretch}{1.6}
\begin{tabular}{ | c || c | }
  \hline                       
  $\psi$ & $W_{\psi}(\chi_{3/2},a)$ \\ \hline \hline
 $(m,0,0,0,0,0,0,0,0,0,0)$ & $v_3^2 v_1^{-1}B_{3/2,m}\left(v_1^2v_3^{-1}\right)$ \\ \hline
 $(0,m,0,0,0,0,0,0,0,0,0)$ & $\frac{v_2^2\tilde{B}_{0,m}(v_2^2v_4^{-1})}{\xi(3)}$ \\ \hline
 $(0,0,m,0,0,0,0,0,0,0,0)$ & $\frac{\xi (2) v_4 B_{1,m}\left( v_3^2 v_1^{-1}v_4^{-1}\right)}{\xi (3)}$ \\ \hline
 $(0,0,0,m,0,0,0,0,0,0,0)$ & $\frac{v_4\tilde{B}_{1/2,m}(v_4^2v_2^{-1}v_3^{-1}v_5^{-1})}{\xi(3)}$  \\ \hline
 $(0,0,0,0,m,0,0,0,0,0,0)$ & $\frac{v_5^2\tilde{B}_{0,m}(v_5^2v_4^{-1}v_6^{-1})}{\xi(3)v_6}$ \\ \hline
 $(0,0,0,0,0,m,0,0,0,0,0)$ & $\frac{\xi (2) v_6^3 B_{-1/2,m}\left(v_6^2 v_5^{-1}v_7^{-1}\right)}{\xi (3)v_7^2}$  \\ \hline
 $(0,0,0,0,0,0,m,0,0,0,0)$ & $ v_7^4v_8^{-3} B_{-1,m}\left(v_7^2 v_6^{-1}v_8^{-1}\right)$  \\ \hline
 $(0,0,0,0,0,0,0,m,0,0,0)$ & $\frac{\xi (4) v_8^5v_9^{-4} B_{-3/2,m}\left(v_8^2 v_7^{-1}v_9^{-1}\right)}{\xi (3)}$  \\ \hline
 $(0,0,0,0,0,0,0,0,m,0,0)$ & $\frac{\xi (5) v_9^6v_{10}^{-5} B_{-2,m}\left(v_9^2 v_8^{-1}v_{10}^{-1}\right)}{\xi (3)}$  \\ \hline
 $(0,0,0,0,0,0,0,0,0,m,0)$ & $\frac{\xi (6) v_{10}^7v_{11}^{-6} B_{-5/2,m}\left(v_{10}^2 v_9^{-1}v_{11}^{-1}\right)}{\xi (3)}$  \\ \hline
  $(0,0,0,0,0,0,0,0,0,0,m)$ & $\frac{\xi (7) v_{11}^8 B_{-3,m}\left(v_{11}^2 v_{10}^{-1}\right)}{\xi (3)}$  \\ \hline
 \end{tabular}
\end{center}

\subsection{$E_{10}$ with $s=5/2$ or $\partial^4\mathcal{R}^4$}
\label{sec:E1052}

For the case $s=\frac52$ we only give exemplary expressions for some degenerate Whittaker vectors for the case of $E_{10}$. Here there are now two types of degenerate Whittaker vectors that are non-zero. The first is of $A_1$-type and the second is of $A_1\times A_1$-type. They correspond to the cases when the degenerate character $\psi$ has support either on a single node or on two disconnected nodes of the Dynkin diagram,

The expressions for the $A_1$-type degenerate Whittaker vectors are much longer now and we do not give all of them. An illustrative example is obtained for instanton charge vector $(m,0,0,0,0,0,0,0,0,0)$. There we have
\begin{align}
W_{\psi}(\chi_{5/2},a) &=\frac{v_3^4   B_{\frac{5}{2},m}\left(\frac{v_1^2}{v_3}\right)}{v_1^3}
 +\frac{\xi (2) v_1^3 v_5^2 B_{-\frac{1}{2},m}\left(\frac{v_1^2}{v_3}\right)}{\xi (5) v_4}
+\frac{\xi (2) v_1^3 v_7^2 B_{-\frac{1}{2},m}\left(\frac{v_1^2}{v_3}\right)}{\xi (5) v_8}+\nn\\
& +\frac{\xi (4)  v_1^3 v_2^4  B_{-\frac{1}{2},m}\left(\frac{v_1^2}{v_3}\right)}{\xi (5) v_3^2}
  +\frac{\xi (3)  v_1^3 v_4^3  B_{-\frac{1}{2},m}\left(\frac{v_1^2}{v_3}\right)}{\xi (5) v_2^2 v_3^2}+v_1^3 v_{10}^5 B_{-\frac{1}{2},m}\left(\frac{v_1^2}{v_3}\right)\nn\\
&+\frac{\xi (3)  v_1^3 v_8^3  B_{-\frac{1}{2},m}\left(\frac{v_1^2}{v_3}\right)}{\xi (5) v_9^2}+\frac{\xi (4)  v_1^3 v_9^4  B_{-\frac{1}{2},m}\left(\frac{v_1^2}{v_3}\right)}{\xi (5) v_{10}^3}\nn\\
&+\frac{ (\gamma_\text{E}-\log(4 \pi  )+ \log(v_5^{-1}v_6^2v_7^{-1})) v_1^3 v_6 B_{-\frac{1}{2},m}\left(\frac{v_1^2}{v_3}\right)}{\xi(5)}.
\end{align}
Here, $\gamma_E\approx 0.577216$ denotes the Euler--Mascheroni constant.

Some non-vanishing degenerate Whittaker vectors with two non-vanishing charges are given by
\begin{center}
\renewcommand{\arraystretch}{2.6}
\begin{tabular}{ | c || c | }
  \hline                       
  $\psi$ & $W_{\psi}(\chi_{5/2},a)$ \\ \hline \hline
 $(m_1,m_2,0,0,0,0,0,0,0,0)$ & $\frac{\xi (3)  v_1^3 v_2^4 B_{-1/2,m_1}\left(\frac{v_1^2}{v_3}\right)B_{-1,m_2}\left(\frac{v_2^2}{v_4}\right)}{\xi (5) v_3^2}$\\\hline
 $(m_1,0,0,m_2,0,0,0,0,0,0)$ & $ \frac{\xi (2) v_1^3 v_4^3 B_{-1/2,m_1}\left(\frac{v_1^2}{v_3}\right)B_{-1/2,m_2}\left(\frac{v_4^2}{v_2v_3v_5}\right)}{\xi (5) v_2^2 v_3^2}$\\\hline
 $(m_1,0,0,0,0,m_2,0,0,0,0)$ & $ \frac{  v_1^3 v_6 B_{-1/2,m_1}\left(\frac{v_1^2}{v_3}\right)\tilde{B}_{1/2,m_2}\left(\frac{v_6^2}{v_5v_7}\right)}{ \xi (5)}$\\\hline
 $(m_1,0,0,0,0,0,0,m_2,0,0)$ & $ \frac{\xi (2) v_1^3 v_8^3 B_{-1/2,m_1}\left(\frac{v_1^2}{v_3}\right)B_{-1/2,m_2}\left(\frac{v_8^2}{v_7v_9}\right)}{\xi (5) v_9^2}$\\\hline
 $(m_1,0,0,0,0,0,0,0,m_2,0)$ & $ \frac{\xi (3)  v_1^3 v_9^4 B_{-1/2,m_1}\left(\frac{v_1^2}{v_3}\right) B_{-1,m_2}\left(\frac{v_9^2}{v_8v_{10}}\right)}{ \xi (5) v_{10}^3}$\\\hline
 $(m_1,0,0,0,0,0,0,0,0,m_2)$ & $ \frac{\xi (4) v_1^3 v_{10}^5 B_{-1/2,m_1}\left(\frac{v_1^2}{v_3}\right)B_{-3/2,m_2}\left(\frac{v_{10}^2}{v_9}\right)}{ \xi (5)}$\\\hline
\ldots&\ldots\\\hline
 \end{tabular}
\end{center}

\newpage

%%%%%%%%%%%
\section{Conclusions}
\label{sec_conclusions}

In this paper we have analyzed Fourier coefficients of Eisenstein series on Kac--Moody groups. We showed that for special points in the parameter space of these Eisenstein series the 
number of such coefficients is drastically reduced, as was previously observed for the constant terms \cite{FK2012}. The non-vanishing coefficients are given by Whittaker vectors associated with  degenerate characters, and for $E_9$, $E_{10}$ and $E_{11}$ we showed that these can be written as linear combinations of simple Bessel-type Whittaker vectors for $A_1=SL(2,\mathbb{R})$ or $A_1\times A_1$. Physically, these Fourier coefficients capture instanton effects in certain string theory amplitudes. As discussed in detail in \cite{Pioline:2010kb,GreenSmallRep,FK2012}, different types of instanton effects correspond to Fourier expansions with respect to different parabolic subgroups. In the case of string theory amplitudes in $D\leq 2$ dimensions (which is the relevant situation where $E_9, E_{10}, E_{11}$-symmetries appear) it is not well-understood precisely which instanton effects contribute, although there are  expectations that they will contain ``exotic'' effects with unusual exponential dependences on the couplings \cite{Blau:1997du,ObersUdualMTheory,deBoer:2012ma}. (For $D\geq 3$, these states were classified in~\cite{Kleinschmidt:2011vu}.) Moreover, BPS-states in $D=2$  fill out infinite-dimensional orbits of the duality group \cite{Englert:2007qb}, in contrast to the case of $D\geq 3$. For a given Fourier coefficient, there is however only a finite number of instanton configurations that contribute to 1/2 and 1/4 BPS-saturated couplings in $D\leq 2$. It would be interesting to examine in more detail how our degenerate Whittaker vectors contribute to the various relevant physical limits~\cite{Green:2010wi,Pioline:2010kb,GreenSmallRep}, as done for the constant terms in \cite{FK2012}. In particular, from the expansions of the Bessel functions one could extract the exponential dependence of the physical couplings in the various limits, thereby giving an indication of which instanton effects are captured by the Eisenstein series. We hope to return to these issues in future work.

It is  natural to speculate about the representation-theoretic implications of our results. As mentioned in the introduction, for the finite-dimensional Lie groups $E_{n}(\mathbb{R})$ ($n=1, \dots, 8$) in Table~\ref{Udualitygroups} the Eisenstein series $E(2s\Lambda_1-\rho, g)$ is attached to certain special unitary representations with unusually small functional dimension. More precisely, for $s=3/2$ this gives an automorphic realization of the minimal representation \cite{GRS,MR0404366,MR1159103,Gunaydin:2001bt,Kazhdan:2001nx}, which is the one with smallest functional dimension (aside from the trivial one), while $s=5/2$ corresponds to the next-to-minimal representation. This means that the character variety orbits on which the Fourier coefficients have support are described by wave-front sets corresponding, respectively, to the closures of the minimal, and next-to-minimal coadjoint nilpotent orbits of $E_{n+1}$ (see \cite{GreenSmallRep,MillerSahi} for more information on these concepts). In the classification of nilpotent orbits of exceptional groups the minimal orbit has Bala--Carter label $A_1$ while the next-to-minimal has label $2A_1$ (see \cite{CollingwoodMcGovern}). Physically, these nilpotent orbits correspond to orbits of the BPS-instantons that contribute to the respective amplitude. 

In the case of Kac--Moody groups it is not known whether there exists some analogue notion of small automorphic representations. The generalization from finite-dimensional groups however, suggests that the values $s=3/2$and $s=5/2$ should play a special role and our investigations corroborate this hypothesis. More rigorously, one would like to have some concept of ``minimal'' and ``next-to-minimal'' nilpotent orbits in this setting, but it is far from clear how to properly define this. Physically we expect that this must be possible, and our results do, indeed, provide some cause for optimism. First of all, we have found that the Kac--Moody Eisenstein series $E(2s\Lambda_1-\rho, g)$ for $E_9$, $E_{10}$ and $E_{11}$ have very few non-vanishing degenerate Whittaker vectors, a feature which is characteristic for small automorphic representations. Moreover, all the non-vanishing Whittaker vectors are associated with subgroups of type $A_1$ or $A_1\times A_1$. By analogy with the finite-dimensional situation, this would suggest that the Eisenstein series with $s=3/2$ has a wave-front set corresponding to the minimal nilpotent orbit of Bala--Carter type $A_1$, and is therefore attached to  what should be called the minimal representation of $E_{9}$ and $E_{10}$. As an example, we note that in the $E_{10}$ case, the weighted Dynkin diagram of the dual (sub-regular) orbit is $[2,2,2,0,2,2,2,2,2,2]$, which has precisely the same structure as for the finite-dimensional groups in the $E_{n}$-series (see \cite{Pioline:2010kb,GreenSmallRep}). Similarly, we note that all the non-vanishing Whittaker vectors for the $s=5/2$ Eisenstein series of $E_{10}$ are of type $A_1\times A_1$ which is consistent with the conjecture that the wave-front set in this case is the next-to-minimal orbit, corresponding to a nilpotent orbit with Bala--Carter label $2A_1$. The weighted Dynkin diagram of the dual (sub-sub-regular) nilpotent orbit is $[2,2,2,0,2,0,2,2,2,2]$, which matches with what one would expect from the next-to-minimal automorphic representation. We give a list of distinguished $E_{10}$ gradings in appendix~\ref{app:e10dist}. It would be desirable to have a complete classification of nilpotent orbits for Kac--Moody algebras.

We leave a more careful mathematical analysis of these issues to future work. 

\subsection*{Acknowledgements}
We are  grateful to Henrik Gustafsson for discussions and for collaboration on the closely related work \cite{FGKP}. We also thank Michael Green, Steve Miller, Boris Pioline, Martin Raum, Gordan Savin and Pierre Vanhove for helpful discussions and correspondence. Some of our results have been presented at the workshop ``Whittaker functions: Number Theory, Geometry and Physics'' at Banff, Canada, and A.K. thanks the organizers Ben Brubaker, Daniel Bump, Gautam Chinta, Solomon Friedberg and Paul Gunnells for a very enjoyable workshop and the participants for inspiring  discussions. A.K. and D.P. gratefully acknowledge Chalmers University of Technology and the Albert Einstein Institute in Golm, respectively, for their warm and generous hospitality while part of this work was carried out.  
The work of P.F. is supported by the Erasmus Mundus Joint Doctorate Program by Grant Number 2010-1816 from the EACEA of the European Commission and the Universit\'e de Nice-Sophia Antipolis.

%%%%%%%%%%%%%%%% APPENDICES

\appendix

\section{Examples for finite-dimensional groups}
\label{sec_finite}

In this appendix, we apply the formula (\ref{degW1}) to some finite-dimensional cases of physical interest. These are associated with the groups $E_{6}$, $E_{7}$ and $E_{8}$ and the particular choices of character $\chi$ that arises in string theory, see e.g.~\cite{Green:2010kv}. The Eisenstein series are again maximal parabolic with character $\chi$ determined by the weight
\begin{align}
\lambda= 2s\Lambda_1-\rho,
\end{align}
where $\Lambda_1$ is the fundamental weight of the node labelled $1$ in the Bourbaki convention, cf. figure~\ref{fig:dynkin}. The cases of relevance for half-BPS and quarter-BPS corrections to the four-graviton scattering amplitude are given by the values $s=3/2$ and $s=5/2$, respectively, and correspond to the minimal and next-to-minimal automorphic representation of the groups.

We give the results for $s=3/2$ only and denote the associated character as $\chi_{3/2}$, but of course the method is applicable to any value of $s$. (The resulting expressions just get longer.) For $s=3/2$ our expressions constitute the complete list of non-vanishing Whittaker vectors.

\subsection{$E_6$}

\begin{center}
\renewcommand{\arraystretch}{1.9}
\begin{tabular}{ | c || c | }
  \hline                       
  $\psi$ & $W_{\psi}(\chi_{3/2},a)$ \\ \hline \hline
 $(m,0,0,0,0,0)$ & $v_3^2 v_1^{-1}B_{3/2,m}\left(v_1^2v_3^{-1}\right)$ \\ \hline
 $(0,m,0,0,0,0)$ & $\frac{v_2^2\tilde{B}_{0,m}(v_2^2v_4^{-1})}{\xi(3)}$ \\ \hline
 $(0,0,m,0,0,0)$ & $\frac{\xi (2) v_4 B_{1,m}\left( v_3^2 v_1^{-1}v_4^{-1}\right)}{\xi (3)}$ \\ \hline
 $(0,0,0,m,0,0)$ & $\frac{v_4\tilde{B}_{1/2,m}(v_4^2v_2^{-1}v_3^{-1}v_5^{-1})}{\xi(3)}$  \\ \hline
 $(0,0,0,0,m,0)$ & $\frac{v_5^2\tilde{B}_{0,m}(v_5^2v_4^{-1}v_6^{-1})}{\xi(3)v_6}$ \\ \hline
 $(0,0,0,0,0,m)$ & $\frac{\xi (2) v_6^3 B_{-1/2,m}\left(v_6^2 v_5^{-1}\right)}{\xi (3)}$  \\ \hline
 \end{tabular}
\end{center}

\subsection{$E_7$}

\begin{center}
\renewcommand{\arraystretch}{1.9}
\begin{tabular}{ | c || c | }
  \hline                       
  $\psi$ & $W_{\psi}(\chi_{3/2},a)$ \\ \hline \hline
 $(m,0,0,0,0,0,0)$ & $v_3^2 v_1^{-1}B_{\frac{3}{2},m}\left(v_1^2 v_3^{-1}\right)$ \\ \hline
 $(0,m,0,0,0,0,0)$ & $\frac{v_2^2\tilde{B}_{0,m}(v_2^2v_4^{-1})}{\xi(3)}$ \\ \hline
 $(0,0,m,0,0,0,0)$ & $\frac{\xi (2) v_4 B_{1,m}\left(v_3^2 v_1^{-1}v_4^{-1}\right)}{\xi (3)}$ \\ \hline
 $(0,0,0,m,0,0,0)$ & $\frac{v_4\tilde{B}_{1/2,m}(v_4^2v_2^{-1}v_3^{-1}v_5^{-1})}{\xi(3)}$  \\ \hline
 $(0,0,0,0,m,0,0)$ & $\frac{v_5^2\tilde{B}_{0,m}(v_5^2v_4^{-1}v_6^{-1})}{\xi(3)v_6}$ \\ \hline
 $(0,0,0,0,0,m,0)$ & $\frac{\xi (2) v_6^3 v_7^{-2} B_{-1/2,m}\left(v_6^2 v_5^{-1}v_7^{-1}\right)}{\xi (3)}$  \\ \hline
 $(0,0,0,0,0,0,m)$ & $v_7^4 B_{-1,m}\left(v_7^2 v_6^{-1}\right)$ \\ \hline
 \end{tabular}
\end{center}

\subsection{$E_8$}

\begin{center}
\renewcommand{\arraystretch}{1.9}
\begin{tabular}{ | c || c | }
  \hline                       
  $\psi$ & $W_{\psi}(\chi_{3/2},a)$ \\ \hline \hline
 $(m,0,0,0,0,0,0,0)$ & $v_3^2v_1^{-1} B_{3/2,m}\left(v_1^2v_3^{-1}\right)$ \\ \hline
 $(0,m,0,0,0,0,0,0)$ & $\frac{v_2^2\tilde{B}_{0,m}(v_2^2v_4^{-1})}{\xi(3)}$ \\ \hline
 $(0,0,m,0,0,0,0,0)$ & $\frac{\xi (2) v_4 B_{1,m}\left(v_3^2 v_1^{-1}v_4^{-1}\right)}{\xi (3)}$ \\ \hline
 $(0,0,0,m,0,0,0,0)$ & $\frac{v_4\tilde{B}_{1/2,m}(v_4^2v_2^{-1}v_3^{-1}v_5^{-1})}{\xi(3)}$  \\ \hline
 $(0,0,0,0,m,0,0,0)$ & $\frac{v_5^2\tilde{B}_{0,m}(v_5^2v_4^{-1}v_6^{-1})}{\xi(3)v_6}$ \\ \hline
 $(0,0,0,0,0,m,0,0)$ & $\frac{\xi (2) v_6^3v_7^{-2} B_{-1/2,m}\left(v_6^2 v_5^{-1}v_7^{-1}\right)}{\xi (3)}$  \\ \hline
 $(0,0,0,0,0,0,m,0)$ & $\frac{v_7^4 B_{-1,m}\left(v_7^2 v_6^{-1}v_8^{-1}\right)}{r_8^3}$ \\ \hline
  $(0,0,0,0,0,0,0,m)$ & $\frac{\xi (4) v_8^5 B_{-3/2,m}\left(v_8^2 v_7^{-1}\right)}{\xi (3)}$ \\ \hline
 \end{tabular}
\end{center}

\section{Proof of reduction formula}
\label{sec_proof}
In this appendix, we present an alternative derivation of the reduction formula~\eqref{degW1}, based on a Chevalley basis decomposition of the argument of the character $\chi$ in integral~\eqref{Fw}. 

As in section~\ref{degfour}, we denote by $\psi$ a degenerate character on $N$ that has support along the subset of simple roots $\Pi'\subset \Pi$.

\subsection{The $n=n_cn'$ factorisation}
\label{appN}

For a reduced Weyl word $w$ of length $\ell$, we start with the integral (\ref{Fw}):
\begin{align}
\mathcal{F}_{w,\psi}(\chi)=\int\limits_{N^w_{\{\gamma\}}}\chi(w n_\gamma )\overline{\psi(n_\gamma)}\mathrm{d}n_\gamma\,.
\end{align}
In the first step towards evaluating this integral we use the Chevalley basis notation~\eqref{chevnot} to write elements $n\in N^w_{\{\gamma\}}(\mathbb{A})$ as
\begin{align}
n_\gamma = x_{\gamma_1}(u_1)\cdots x_{\gamma_\ell}(u_\ell).
\end{align}
The positive roots $\gamma_i$ are parametrised in appendix~\ref{param} and satisfy $w\gamma<0$ and in this appendix we will denote these roots by $\{\gamma\}_w$, where the subscript indicates the fixed Weyl word $w$.
We have also used the Chevalley generator notation $x_{\alpha}(u)$ 
\begin{align}
x_\alpha(u) = e^{u E_\alpha}\,,
\end{align}
where $E_\alpha$ is the generator of the $\alpha$ root space and $u\in\mathbb{A}$ is the parameter of the group element. With this parametrisation, we can rewrite the individual term $\mathcal{F}_{w,\psi}$ as
\begin{align}\label{FourChev}
\mathcal{F}_{w,\psi}(\chi)&=\int\limits_{\mathbb{A}^\ell}\chi\left(wx_{\gamma_1}(u_1)...x_{\gamma_\ell}(u_\ell)\right)\overline{\psi(u)}\mathrm{d}u_1...\mathrm{d}u_\ell \nn\\
&=\int\limits_{\mathbb{A}^\ell}\chi\left(x_{w\gamma_1}(u_1)...x_{w\gamma_\ell}(u_\ell)\right)\overline{\psi(u)}\mathrm{d}u_1...\mathrm{d}u_\ell\,,
\end{align}
where $u$ now denotes collectively all $u_i$ variables.\\

Let us label the simple roots of $\Pi'$ more explicitly as
\begin{align}
\Pi'=\{\alpha'_{i_1},...,\alpha'_{i_r}\}
\end{align}
hence $\psi$ is non-trivial along $r$ ``directions''. Since $\psi$ is only sensitive to those components of $n_\gamma$, which correspond to the non-trivial directions, $\Pi'$, we can write
\begin{align}
\psi(u)=\psi(u_{\alpha'_{i_1}},...,u_{\alpha'_{i_r}})\,.
\end{align}
The particular order of the arguments in $\psi$ is not important, since the character is abelian. In the following we will denote the variables $u_{\alpha'_{i_1}}$ to $u_{\alpha'_{i_r}}$ collectively by $u_{\Pi'}$. Let us recall that for a given choice of the set $\Pi'$, one is restricted to the set of Weyl words $\mathcal{C}_\psi$, cf.~\eqref{specialw}. As a consequence, the subset of simple roots $\Pi'$ is contained in the set of roots $\{\gamma\}_w$. The set $\{\gamma\}_w$ provides a canonical ordering of the Chevalley factors in the argument of the character $\chi$ in~\eqref{FourChev}. The embedding of $\Pi'$ in $\{\gamma\}_w$ in general takes the form
\begin{align}
\{\gamma_1,...,\gamma_k, \gamma_{k+1}=\alpha'_{i_1}, ...,\gamma_\ell=\alpha'_{i_r}\}_w\,,
\end{align}
where we have used the fact that all other simple roots of $\Pi'$ appear somewhere to the right of $\gamma_{k+1}$. 
In other words the roots $\{\gamma_1,...,\gamma_k\}$ to the left of $\alpha'_{i_1}$ do not contain simple roots along which $\psi$ is non-trivial. We define 
\begin{align}
\{\gamma\}_{\wlong'}=\{\gamma_{k+1}=\alpha_{i_1}',...,\gamma_\ell=\alpha_{i_r}'\}_{\wlong'}\,.
\end{align}
We will exploit this fact in the following to show that the Fourier integral can be factorized into two integrals, one which is independent of the Fourier kernel $\psi$ and another one which depends on it. We present a constructive proof of this statement.

\subsection{Iterative step}
\label{appIter}

Starting with the integral of~\eqref{FourChev}, we perform an Iwasawa decomposition of the last Chevalley factor $x_{w\gamma_\ell}$ in the argument according to
\begin{align}
x_{w\gamma_\ell}(u_\ell)=n(u_\ell)a(u_\ell)k(u_\ell)\,.
\end{align}
Due to the right-invariance of $\chi$ under the action of elements of $K(\mathbb A)$, we can drop the $k$ factor, leaving us with
\begin{align}
\mathcal{F}_{w,\psi}(\chi)=\int\limits_{\mathbb{A}^\ell}\chi\left(x_{w\gamma_1}(u_1)...x_{w\gamma_{\ell-1}}(u_{\ell-1})n(u_\ell)a(u_\ell)\right)\overline{\psi(u_{\Pi'})}\mathrm{d}u_1...\mathrm{d}u_\ell\,.
\end{align}
We will now move the factor $n(u_\ell)a(u_\ell)$ all the way to the left in the argument of $\chi$, by conjugating it around each one of the other $\ell-1$ remaining Chevalley factors. This process will transform the arguments of each one of these Chevalley factors. Let us first consider the effect of ``pulling through'' the $n(u_\ell)$ factor. The argument of a Chevalley factor will then transform as
\begin{align}
x_{\gamma_i}(u_i)\rightarrow x_{\gamma_i}(u_i+f_i(u_{i+1},...,u_\ell))\,,
\end{align}
where $i=1,...,\ell-1$ and $f_i$ is some polynomial function in its arguments. By making a re-definition of variables according to $u_i \rightarrow u_i-f_i(u_{i+1},...,u_\ell)$, we can restore the simple form in the argument of the Chevalley factor. The important point now is that the corresponding Jacobi factor of this variable transformation is trivial, i.e. equal to one since $u_i$ is shifted only by higher $u_k$, see e.g.~\cite{Liu}. Nevertheless the arguments $u_{\Pi'}$ of the character $\psi$ will obtain a dependence on the variables of Chevalley factors associated with the roots $\{\gamma\}_{\wlong'}$. This dependence is indicated by $u_{\{\gamma\}_{\wlong'}}$. We are are then left with an integral of the form
\begin{align}
\mathcal{F}_{w,\psi}(\chi)=\int\limits_{\mathbb{A}^\ell}\chi\left(n(u_\ell)x_{w\gamma_1}(u_1)...x_{w\gamma_{\ell-1}}(u_{\ell-1})a(u_\ell)\right)\overline{\psi(u_{\{\gamma\}_{\wlong'}})}\mathrm{d}u_1...\mathrm{d}u_\ell\,.
\end{align}
Now consider moving the factor $a(u_\ell)$ to the left by conjugating it around each Chevalley factor. This induces a scaling of the argument of each Chevalley factor according to
\begin{align}
x_{w\gamma_i}(u_i)\rightarrow x_{w\gamma_i}(u_ia(u_\ell)^{-w\gamma_i})\,.
\end{align}
We can make a variable transformation $u_i\rightarrow u_ia(u_\ell)^{w\gamma_i}$ in order to eliminate the scaling from the argument. This transformation yields a Jacobi factor of $J_i(u_\ell)=a(u_\ell)^{w\gamma_i}$. There will of course also be a re-scaling in the argument of the character $\psi$, which we will indicate by a tilde. Following this procedure, we are left with the following expression for the integral
\begin{align}
\mathcal{F}_{w,\psi}(\chi)=&\int\limits_{\mathbb{A}^\ell}\Big(\prod_{i=1}^{\ell-1}J_i(u_\ell)\Big)\chi\left(n(u_\ell)a(u_\ell)x_{w\gamma_1}(u_1)...x_{w\gamma_{\ell-1}}(u_{\ell-1})\right)\nn\\
&\overline{\psi(\tilde{u}_{\{\gamma\}_{\wlong'}})}\mathrm{d}u_1...\mathrm{d}u_\ell\,.
\end{align}
Let us denote the overall Jacobi factor by
\begin{align}\label{jacobiell}
\mathcal{J}_\ell(u_\ell)=\prod_{i=1}^{\ell-1}J_i(u_\ell)
\end{align}
By the definition of the character $\chi$ one can drop the $n(u_\ell)$ factor from its argument. After furthermore splitting off the factor $\chi(a(u_\ell))$, the integral takes the following form
\begin{align}
\mathcal{F}_{w,\psi}(\chi)=&\int\limits_{\mathbb{A}^\ell}\mathcal{J}_\ell(u_\ell)\,\chi(a(u_\ell))
\chi\left(x_{w\gamma_1}(u_1)...x_{w\gamma_{\ell-1}}(u_{\ell-1})\right)\nn\\
&\overline{\psi(\tilde{u}_{\{\gamma\}_{\wlong'}})}\mathrm{d}u_1...\mathrm{d}u_\ell\,.
\end{align}

\subsection{Iteration and projection}\label{appIterProj}

The procedure outlined in the previous section is applied in an iterative process to each Chevalley factor in the argument of $\chi$. An important step is the re-scaling transformation used to undo the effect that the conjugate action of an abelian element $a(u_i)$ has on the argument of a Chevalley factor. The point is that \textit{only} the re-scaling in the Chevalley factors $x_{\{\gamma\}_{\wlong'}}$ will also have an effect on the variables on which the Fourier kernel $\psi$ depends. To see this more clearly, let us write the integral~\eqref{FourChev} for $\mathcal{F}_{w,\psi}$ in the following shorthand notation
\begin{align}
\mathcal{F}_{w,\psi}(\chi)&=\int\limits_{\mathbb{A}^\ell}\chi\left(x_{w(\{\gamma\}_w\backslash\{\gamma\}_{\wlong'})}\;x_{w(\{\gamma\}_{\wlong'})}(u)\right)\overline{\psi(u_{\Pi'})}\mathrm{d}u\,,
\end{align}
where we have defined the following notation
\begin{align}
\{\gamma\}_w\backslash\{\gamma\}_{\wlong'}:=\{\gamma_1,...,\gamma_k\}\,.
\end{align}
Having performed the iteration step of~\ref{appIter} on each Chevalley factor $x_{w(\{\gamma\}_{\wlong'})}$ successively and defining a generalised analogue of the Jacobi factor~\eqref{jacobiell} according to
\begin{align}
\mathcal{J}_i(u_i)=\prod_{j=1}^{i-1}J_j(u_i)\,,
\end{align}
we obtain the following structure of the integral
\begin{align}
\mathcal{F}_{w,\psi}(\chi)=\int\limits_{\mathbb{A}^\ell}\chi\left(x_{w(\{\gamma\}_w\backslash\{\gamma\}_{\wlong'})}\right)
\left(\prod_{i=k+1}^{k+\ell'}\mathcal{J}_i(u_i)\,\chi(a(u_i))\right)
\overline{\psi(\tilde{u}_{\{\gamma\}_{\wlong'}})}\mathrm{d}u\,,
\end{align}
where $\ell'=\ell(\wlong')$. It is now clear that we can split the integral up into a product of one integral which is independent of $\psi$ and a second one which retains dependence on $\psi$
\begin{align}
\mathcal{F}_{w,\psi}(\chi)&=\int\limits_{\mathbb{A}^{\ell-\ell'}}\chi\left(x_{w(\{\gamma\}_w\backslash\{\gamma\}_{\wlong'})}\right)\mathrm{d}u_1...\mathrm{d}u_{\ell-\ell'}\nn\\
&\cdot\int\limits_{\mathbb{A}^{\ell'}}
\left(\prod_{i=k+1}^{k+\ell'}\mathcal{J}_i(u_i)\,\chi(a(u_i))\right)
\overline{\psi(\tilde{u}_{\{\gamma\}_{\wlong'}})}\mathrm{d}u_{\ell-\ell'+1}...\mathrm{d}u_\ell\,.
\end{align}
Let us denote the factors as follows
\begin{align}\label{Pwlong}
\mathcal{P}_w&=\int\limits_{\mathbb{A}^{\ell-\ell'}}\chi\left(x_{w(\{\gamma\}_w\backslash\{\gamma\}_{\wlong'})}\right)\mathrm{d}u_1...\mathrm{d}u_{\ell-\ell'}\,,\nn\\
\mathcal{P}_{\wlong',\psi}&=\int\limits_{\mathbb{A}^{\ell'}}
\left(\prod_{i=k+1}^{k+\ell'}\mathcal{J}_i(u_i)\,\chi(a(u_i))\right)
\overline{\psi(\tilde{u}_{\{\gamma\}_{\wlong'}})}\mathrm{d}u_{\ell-\ell'+1}...\mathrm{d}u_\ell\,,
\end{align}
such that $\mathcal{F}_{w,\psi}=\mathcal{P}_w\cdot\mathcal{P}_{\wlong',\psi}$. Written out with the character $\chi$ parametrised by $\lambda$, we have
\begin{align}
\mathcal{J}_i(u_i)\,\chi(a(u_i))=a(u_i)^{\lambda+\rho+w(\sum_{j=1}^{i-1}\gamma_j)}\,.
\end{align}

As the next step, we will now show that the integral $\mathcal{P}_{\wlong',\psi}$ can be conveniently projected onto the subgroup $G'\subset G$ generated by the simple roots in $\Pi'$. To achieve this, we define an Eisenstein series $E^{G'}(\chi',g')$ on the group $G'$. Since $\psi$ is a \textit{generic} character on $G'$, the projection of the integral $\mathcal{P}_{\wlong',\psi}$ then gives the \textit{non-degenerate} Whittaker vector of the $E^{G'}(\chi',g')$ series. This fact allows one to apply the Shintani--Casselman-Shalika formula~\cite{Shintani,CasselmanShalika} for evaluating the integral.\\

The non-degenerate Whittaker vector $W'_{\psi}(\chi',g')$, associated with the series $E^{G'}(\chi',g')$, and evaluated at $g'=a'=\id$, takes the form
\begin{align}
W'_{\psi'}(\chi',\id)&=\int\limits_{N'(\mathbb{Q})\backslash N'(\mathbb{A})}\chi'(\wlong'n')\overline{\psi(n')}
\end{align}
In the following we use
\begin{align}
\{\gamma'\}_{\wlong'}=\{\gamma'_1,...,\gamma'_{\ell'}\}\,.
\end{align}
With this, the integral for $W'_{\psi}(\chi',\id)$ evaluates to
\begin{align}\label{Wwlong}
\int_{\mathbb{A}^{\ell'}}\left(\prod_{i=1}^{\ell'}\mathcal{J}_i(u_i)\chi'(a(u_i))\right)\overline{\psi(u_{\{\gamma'\}_{\wlong'}})}\mathrm{d}u_{1}...\mathrm{d}u_{\ell'}
\end{align}
In order to project~\eqref{Pwlong} onto~\eqref{Wwlong}, we have to identify the two Fourier integrands with each other. This yields the condition for the exponents
\begin{align}\label{equivcalc1}
\langle\lambda'+\rho'+\wlong'(\sum_{j=k+1}^{i-1}\gamma_j)|H_{-\wlong'\gamma_i}\rangle=\langle\lambda+\rho+w(\sum_{j=1}^{i-1}\gamma_j)|H_{-w\gamma_i}\rangle\,,
\end{align}
where on the right-hand side, we have used the trivial observation that $\{\gamma'\}_{\wlong'}=\{\gamma_{k+1},...,\gamma_{\ell}\}$. For a positive root $\alpha$, we have denoted by $H_\alpha=[E_\alpha,E_{-\alpha}]$ the canonically normalized element in the Cartan subalgebra.

Now recall that $w=w_c\wlong'$, such that the right-hand side can be re-written as
\begin{align}
\langle w_c^{-1}(\lambda+\rho)+\wlong'(\sum_{j=1}^{i-1}\gamma_j)|w_c^{-1}H_{-w\gamma_i}\rangle=\langle w_c^{-1}(\lambda+\rho)+\wlong'(\sum_{j=1}^{i-1}\gamma_j)|H_{-\wlong'\gamma_i}\rangle\,.
\end{align}
With this,~\eqref{equivcalc1} reduces to
\begin{align}
\langle\lambda'+\rho'|H_{-\wlong'\gamma_i}\rangle=\langle w_c^{-1}(\lambda+\rho)+w_c^{-1}w_c\wlong'(\sum_{j=1}^{k}\gamma_j)|H_{-\wlong'\gamma_i}\rangle\,.
\end{align}
Applying the identity~\eqref{wsumgamma}, we obtain
\begin{align}
\langle\lambda'+\rho'|H_{-\wlong'\gamma_i}\rangle&=\langle w_c^{-1}(\lambda+\rho)+w_c^{-1}(w_c\rho-\rho)|H_{-\wlong'\gamma_i}\rangle\nn\\
&=\langle w_c^{-1}\lambda+\rho|H_{-\wlong'\gamma_i}\rangle\,,
\end{align}
giving us the desired parametrisation of $\lambda'$ in terms of $\lambda$.

\section{The $\gamma$-parametrisation}\label{param}

We require a parametrisation of $N^w_{\{\gamma\}}$ in~\eqref{NGamma}. In other words we seek a construction of the set of roots $\{\gamma\}$, for given $w$. Though this construction is standard (see e.g.~\cite[Lemma 1.3.14]{KumarBook}), we give it for completeness and employ a construction obtained in~\cite{Liu}.

For this we fix a reduced expression $w=w_{i_1}w_{i_2}\cdots w_{i_\ell}$ for the Weyl word $w$ of length $\ell$. The subscripts refer to the nodes of the Dynkin diagram of $G$ and $w_i$ are the fundamental reflections that generate the Weyl group. Then one can explicitly enumerate all positive roots that are mapped to negative roots by the action of $w$ as follows. Define
\begin{align}\label{gammak}
\gamma_{k} = w_{i_\ell} w_{i_{\ell-1}}\cdots w_{i_{k+1}}\alpha_{i_k}
\end{align}
where $\alpha_{i_k}$ is the $i_k$th simple root. We note in particular $\gamma_\ell=\alpha_{i_\ell}$. That this gives a valid description of the positive roots generating $N^w_{\{\gamma\}}$ can be checked easily by induction. Therefore we have
\begin{align}
\{\gamma\}_w=\{\alpha>0\,|\, w\alpha<0\} = \{\gamma_i \,:\, i=1,\ldots,\ell(w)\}\,,
\end{align}
where for clarity we have introduced the subscript label $w$.
We also record that there is a simple expression for the sum of all these roots in terms of the Weyl vector $\rho$
\begin{align}
\label{sumgamma}
\gamma_1+\ldots+\gamma_\ell = \rho - w^{-1} \rho
\end{align}
which can again be checked by induction. Furthermore, we record that
\begin{align}\label{wsumgamma}
w(\gamma_1+...+\gamma_n)=w'\rho-\rho\,,
\end{align}
where $w'=w_{i_1}w_{i_2}\cdots w_{i_n}$, with $n<\ell$.

It is important to emphasise that by the construction~\eqref{gammak}, the set inherits a canonical order of the roots, such that $\{\gamma\}_w=\{\gamma_1,\gamma_2,...,\gamma_\ell\}$ denotes an ordered set.

\section{Distinguished gradings of $E_{10}$}
\label{app:e10dist}

For a finite-dimensional (simple, complex) Lie algebra $\mathfrak{g}$ one can study the so-called \textit{distinguished gradings} of $\mathfrak{g}$. These are integers gradings
\begin{align}
\mathfrak{g} = \oplus_{m\in\mathbb{Z}} \mathfrak{g}_m
\end{align}
such that $\dim\,\mathfrak{g}_0=\dim\,\mathfrak{g}_2$. These conditions imply that the grading is even: $\mathfrak{g}_1=\{0\}$~\cite{CollingwoodMcGovern}. For finite-dimensional Lie algebras, the set of all distinguished gradings is in one-to-one correspondence with nilpotent orbits with Bala--Carter labels of type $\mathfrak{g}$~\cite{BalaCarterI,BalaCarterII}.\footnote{On top of these, there are nilpotent orbits associated with distinguished gradings of Levi factors of parabolic subalgebras of $\mathfrak{g}$ and together they provide a complete classification of nilpotent orbits.} The distinguished gradings can be given in terms of weighted Dynkin diagrams $[p_1,\ldots, p_n]$ that determine the degree of any root $\alpha$ by
\begin{align}
m=\mathrm{deg}\,\alpha = \sum_i p_i n_i
\end{align}
if $\alpha=\sum_i n_i \alpha_i$ is expanded on a basis of simple roots $\alpha_i$ with integer coefficients $n_i$. Evenness of the grading means that all $p_i$ are even numbers. For finite-dimensional $\mathfrak{g}$ the maximum value is $p_i=2$~\cite{CollingwoodMcGovern}, so that all $p_i$ are either zero or equal to two. The principal nilpotent orbit corresponds to $[2,2,\ldots,2]$ and is the largest possible orbit with dimension $(\dim\,\mathfrak{g}) - (\mathrm{rk}\,\mathfrak{g})$, i.e. codimension equal to the rank of $\mathfrak{g}$.

\begin{table}[t!]
\centering
\begin{tabular}{p{7mm}p{4mm}p{4mm}p{4mm}p{4mm}p{4mm}p{4mm}p{4mm}p{4mm}p{6mm}||c}
 $[p_1$&$p_2$&.&.&.&.&.&.&$p_9$&$p_{10}]$ & Codimension\\[2mm]\hline\hline
&&&&&&&&&&\\[-3mm]
 [ 0 & 0 & 0 & 0 & 2 & 0 & 0 & 0 & 0 & 0 ]&60\\{}
[ 0 & 0 & 2 & 0 & 0 & 0 & 0 & 0 & 2 & 0 ]&56\\{}
[ 0 & 0 & 2 & 0 & 0 & 0 & 0 & 2 & 0 & 0 ]&48\\{}
[ 0 & 0 & 2 & 0 & 0 & 0 & 0 & 2 & 0 & 2 ]&44\\{}
[ 0 & 0 & 0 & 2 & 0 & 0 & 0 & 0 & 2 & 0 ]&40\\{}
[ 0 & 0 & 2 & 0 & 0 & 0 & 2 & 0 & 0 & 2 ]&38\\{}
[ 0 & 0 & 0 & 0 & 2 & 0 & 0 & 2 & 0 & 2 ]&38\\{}
[ 0 & 0 & 0 & 2 & 0 & 0 & 0 & 2 & 0 & 0 ]&36\\{}
[ 2 & 0 & 0 & 0 & 2 & 0 & 0 & 0 & 2 & 0 ]&36\\{}
[ 0 & 0 & 0 & 2 & 0 & 0 & 2 & 0 & 0 & 2 ]&30\\{}
[ 2 & 0 & 0 & 0 & 2 & 0 & 0 & 2 & 0 & 2 ]&30\\{}
[ 0 & 0 & 0 & 2 & 0 & 0 & 0 & 2 & 2 & 2 ]&30\\{}
[ 0 & 0 & 0 & 2 & 0 & 0 & 2 & 0 & 2 & 0 ]&28\\{}
[ 2 & 0 & 0 & 2 & 0 & 0 & 0 & 2 & 0 & 2 ]&28\\{}
[ 2 & 0 & 0 & 2 & 0 & 0 & 2 & 0 & 0 & 2 ]&26\\{}
[ 2 & 0 & 0 & 2 & 0 & 0 & 2 & 0 & 2 & 0 ]&24\\{}
[ 0 & 0 & 0 & 2 & 0 & 0 & 2 & 2 & 2 & 2 ]&24\\{}
[ 2 & 0 & 0 & 2 & 0 & 2 & 0 & 2 & 0 & 2 ]&20\\{}
[ 2 & 0 & 0 & 2 & 0 & 0 & 2 & 2 & 2 & 2 ]&20\\{}
[ 2 & 0 & 0 & 2 & 0 & 2 & 0 & 2 & 2 & 2 ]&18\\{}
[ 2 & 2 & 2 & 0 & 2 & 0 & 2 & 0 & 2 & 2 ]&16\\{}
[ 2 & 0 & 0 & 2 & 0 & 2 & 2 & 2 & 2 & 2 ]&16\\{}
[ 2 & 2 & 2 & 0 & 2 & 0 & 2 & 2 & 2 & 2 ]&14\\{}
[ 2 & 2 & 2 & 0 & 2 & 2 & 2 & 2 & 2 & 2 ]&12\\{}
[ 2 & 2 & 2 & 2 & 2 & 2 & 2 & 2 & 2 & 2 ]&10
\end{tabular}
\caption{\label{tab:e10dist}\sl Distinguished even gradings of $E_{10}$ with $p_i\leq 2$. The last column contains the potential codimension of the associated nilpotent orbit.}
\end{table}

One can apply the same criterion for distinguished even gradings to Kac--Moody algebras. It appears that the bound $p_i\leq 2$ is no longer satisfied, leading to a much larger (and possibly infinite) set of distinguished gradings. In Table~\ref{tab:e10dist}, we list all distinguished gradings for $E_{10}$ that have $p_i\leq 2$.

The last row of the table corresponds to what is the principal orbit in the finite-dimensional case and should be associated with the generic principal series representation. The next-to-last (sub-regular) and second-to-last (sub-sub-regular) rows were discussed in the conclusions in relation to the minimal and next-to-minimal representations of $E_{10}$. (This also requires the definition of an analogue of the Spaltenstein map~\cite{MR672610} for Kac--Moody algebras.)

{\small 
\bibliography{Whittakerbib}
\bibliographystyle{utphys}
}

\end{document}